\documentclass{jpp}
\usepackage{graphicx}

\usepackage[utf8]{inputenc}
\usepackage[T1]{fontenc}
\usepackage{amsmath}
\usepackage{comment}
\usepackage{url}
\usepackage{hyperref}

\newcommand{\nfp}{n_{\mathrm{fp}}}

\usepackage{color}

\newcommand{\changed}[1]{{#1}}

\shorttitle{How does ITG turbulence depend on magnetic geometry?}
\shortauthor{
$\;\;\;\;$M Landreman, 
J Choi, 
C Alves,
P Balaprakash, 
R Churchill,
R Conlin,
G Roberg-Clark
}

\title{How does ion temperature gradient turbulence depend on magnetic geometry? Insights from data and machine learning}

\author{Matt Landreman\aff{1}
  \corresp{\email{mattland@umd.edu}},
  Jong Youl Choi\aff{2},
  Caio Alves\aff{2},
  Prasanna Balaprakash\aff{2}, 
  R. Michael Churchill\aff{3},
  Rory Conlin\aff{1},
  Gareth Roberg-Clark\aff{4}
 }

\affiliation{\aff{1}Institute for Research in Electronics \& Applied Physics, University of Maryland,
College Park, MD 20742, USA
\aff{2}Oak Ridge National Laboratory, Oak Ridge, TN 37831, USA
\aff{3}Princeton Plasma Physics Laboratory, Princeton, NJ 08540, USA
\aff{4}Max Planck Institute for Plasma Physics, Wendelsteinstraße 1, 17491 Greifswald, Germany
}

\begin{document}

\maketitle

\begin{abstract}
Magnetic geometry has a significant effect on the level of turbulent transport in fusion plasmas.
Here, we model and analyze this dependence using multiple machine learning methods and a dataset of $>2\times 10^5$ nonlinear gyrokinetic simulations of ion-temperature-gradient turbulence in diverse non-axisymmetric geometries.
The dataset is generated using a large collection of both optimized and randomly generated stellarator equilibria.
At fixed gradients and other input parameters, the turbulent heat flux varies between geometries by several orders of magnitude.
Trends are apparent among the configurations with particularly high or particularly low heat flux.
Regression and classification techniques from machine learning are then applied to extract patterns in the dataset.
Due to a symmetry of the gyrokinetic equation, the heat flux and regressions thereof should be invariant to translations of the raw features in the parallel coordinate, similar to translation invariance in computer vision applications.
Multiple regression models including convolutional neural networks (CNNs) and decision trees can achieve reasonable predictive power for the heat flux in held-out test configurations, with highest accuracy for the CNNs.
Using Spearman correlation, sequential feature selection, and Shapley values to measure feature importance, it is consistently found that the most important geometric lever on the heat flux is the flux surface compression in regions of bad curvature.
The second most important geometric feature relates to the magnitude of geodesic curvature.
These two features align remarkably with surrogates that have been proposed based on theory, while the methods here allow a natural extension to more features for increased accuracy.
The dataset, released with this publication, may also be used to test other proposed surrogates, and we find that many previously published proxies do correlate well with both the heat flux and stability boundary.
\end{abstract}


\section{Introduction}

The level of turbulent transport in magnetized plasmas depends on the magnetic field geometry in complicated ways.
In tokamaks,  geometric factors include the aspect ratio, elongation, and triangularity, while in stellarators, there is vast additional freedom in the nonaxisymmetric shaping.
While direct numerical simulation of the turbulence can be used to evaluate the transport for specific geometries, numerical calculations do not immediately give insight into general parametric dependencies. 
Numerical turbulence simulations also require non-negligible computation time and are noisy due to the chaotic dynamics, making it challenging to include them directly in optimization or parametric studies.
Therefore, a variety of physics-based surrogates for the geometry-dependence of turbulent transport have been proposed \changed{\citep{mynick2010optimizing, xanthopoulos2014controlling, mynick2014turbulent, proll2015tem, hegna2018theory, mackenbach2022available,NAKATA2022,nakayama2023simplified, roberg2023critical,goodman2024quasi}}.
However, owing to advances in computing hardware and simulation software, it is now possible to assemble large sets of turbulence simulation data that span a wide range of possible geometries, even in the high-dimensional parameter space of stellarators.
Moreover, progress in machine learning (ML) methods and software provide many new opportunities for understanding and exploiting this data.
The purpose of this article is to show how ML methods can discover patterns in the geometry-dependence of plasma turbulence.
Specifically, we present a new method for generating training data using random 3D geometries, we compare the accuracy of several machine learning methods at predicting the turbulent heat flux, and we show several interpretable ML techniques that can identify which geometric factors determine the turbulent heat flux.
This last aspect shows that ML can be more than a black-box interpolation method to accelerate computations- it can in fact also feed back into more traditional physics analysis.
For example, the patterns in the data identified here can provide evidence for or against physics-inspired approximate models, and motivate future theoretical studies.

The methods here extend prior work in several ways.
ML methods, specifically neural networks, have been applied previously as surrogates for transport in tokamaks \citep{meneghini2014modeling, citrin2015real, meneghini2017self, narita2019neural, honda2019machine, van2020fast, boyer2021prediction, abbate2021data, li2024surrogate}. 
In contrast to this previous work, here we allow for nonaxisymmetric shaping, in which case the geometric parameter space is much higher dimensional.
In the context of stellarators, neural networks have been applied to neoclassical transport coefficients \citep{wakasa2007construction} and magneohydrodynamic (MHD) equilibria \citep{merlo2021proof, curvo2024using}.
For stellarator turbulent transport, regressions have been performed using theory-based analytic models with tuning parameters \citep{nakayama2023simplified}.
Theory-based surrogates for transport in stellarators have previously been compared to gyrokinetic simulations for a small number of geometries $N$, e.g. $N=10$ in \cite{proll2015tem}, $N=4$ in \cite{mackenbach2022available}, or $N=9$ in \cite{roberg2023critical}.
In the present work the number of geometries considered is increased by 4 orders of magnitude to  $N> 10^5$.
Compared to these earlier studies, the work here is also unique in the new method of data generation using random novel 3D geometries, in the application of both neural-network and non-neural-network-based ML methods, and in the use of interpretable ML methods to identify important features in the geometry.
Finally, in supplemental material available at \citep{zenodo}, we are making the training dataset publicly available, so other researchers can test the accuracy of different proposed turbulence surrogates.

In order to obtain ML models for this problem that are interpretable, we demonstrate a novel approach that combines a large library of candidate features with forward sequential feature selection (FSFS) and traditional regression and classification methods.
The new approach here is used to ensure that the models respect a translation-invariance in the gyrokinetic equation: the heat flux should be invariant to periodic translation of all geometric quantities in the direction along the magnetic field (see section \ref{sec:GKE}).
The invariance is guaranteed in our models by using a library of features that respect the symmetry.
The use of a combinatorial library of candidate features in our method is reminiscent of SINDy (sparse identification of nonlinear dynamics) \citep{brunton2016discovering}, except that we are not interested in time-dependence, and we obtain parsimony through FSFS rather than via sparsity-promoting optimization.
Our method is also reminiscent of symbolic regression \citep{koza1994genetic} in that we seek symbolic expressions for interpretability.
However unlike SINDy and symbolic regression, the method here combines the symbolic feature library with additional regression and classification models (decision trees and nearest-neighbors) to efficiently allow extra nonlinearity.

We find a remarkable alignment between the results of interpretable ML analysis here and recently proposed physics-inspired surrogates for turbulence.
Across several ML regression methods and several ways to measure feature importance, the most important two features are consistent.
The most important geometric factor is found to be the flux surface compression $|\nabla\psi|$ in regions of bad curvature, where $2\pi\psi$ is the toroidal flux, reflecting the gradient drive in real space in regions of linear instability.
This factor has been used as an optimization objective function for ion temperature gradient (ITG) turbulence by \cite{mynick2014turbulent}, \cite{xanthopoulos2014controlling}, \cite{stroteich2022seeking}, and \cite{goodman2024quasi}, and a more complicated objective with these elements was used previously by \cite{mynick2010optimizing}.
The second most important geometric factor identified in our analysis is the average magnitude of the geodesic curvature, equivalent to the radial magnetic drift $\propto \mathbf{B}\times\mathbf{\kappa}\cdot\nabla\psi$ where $\mathbf{\kappa}$ is the curvature.
This quantity has been proposed as a correlate of turbulence by \cite{xanthopoulos2011zonal} and \cite{NAKATA2022}.
The theoretical motivation for this quantity is related to zonal flows: larger geodesic curvature leads to stronger linear damping of zonal flows, resulting in \changed{higher levels of saturated heat flux}.
For both of two most important features, our analysis matches the theoretical predictions for the sign of the correlation: increased $|\nabla\psi|$ and increased absolute geodesic curvature correlate with heat flux increase.

Since ML methods are most effective when large amounts of data are available that cover the parameter space of interest, we make several simplifying assumptions in this work.
First, we consider the electrons to be adiabatic.
This choice makes the direct numerical simulations faster by a factor of $\sim \sqrt{m_i / m_e}$, where $m_i$ and $m_e$ are the ion and electron masses.
Hence, the maximum stable time step is not limited by the electron parallel speed, increasing the number of simulations that can be run for a given computational budget.
\changed{Compared to simulations with adiabatic electrons, simulations with kinetic electrons have higher heat fluxes \citep{chen2003simulations}, and the difference depends on magnetic geometry \citep{mckinney2019comparison, goodman2024quasi}, so some conclusions of the analysis here may be modified if revisited with kinetic electrons.}
By the use of adiabatic electrons, our analysis is necessarily electrostatic, and focuses on ion temperature gradient turbulence.

When attempting to model the turbulent heat flux, there are two natural options for the independent geometric variables: one could either use shape parameters of the plasma, or the geometric quantities appearing directly in the gyrokinetic equation. Examples of the former include elongation and triangularity, or Fourier amplitudes of the boundary surface, while examples of the latter include the field magnitude $|B|$ and the guiding center drift component $B^{-3}\mathbf{B}\times\nabla B \cdot\nabla\psi$.
By taking the independent variables to be the plasma shape parameters, an ML regression effectively models physics of both MHD equilibrium and turbulence.
If the independent variables are instead taken to be the geometry factors in the gyrokinetic equation, the ML regression becomes a model only for the turbulence, without including MHD equilibrium physics.
Here we take the latter approach, because by focusing on a narrower aspect of the physics, there is greater hope for interpretability of the model.
At the same time, we nonetheless generate the geometric inputs to the gyrokinetic equation from actual global MHD equilibria so correlations and constraints among the geometric features are respected.

To introduce the methods here in detail, we begin in the following section by reviewing the gyrokinetic equation used in the direct numerical simulations, highlighting the geometric inputs and its translation invariance.
This translation invariance is important because we will want ML models to respect this symmetry. 
The procedure for generating data, including both MHD equilibria and turbulence simulations, is then detailed in section \ref{sec:data_generation}.
Neural network fits to the data are presented in section \ref{sec:CNN}.
Section \ref{sec:features} presents alternative ML methods that do not use neural networks, allowing for greater interpretability at the expense of reduced accuracy.
These alternative methods require manual feature engineering and feature selection.
In section \ref{sec:other_surrogates} it is demonstrated how the same dataset can be used to assess other proposed proxies for turbulence, by evaluating the accuracy of several ITG objective functions from earlier papers.
Finally, we conclude and discuss future directions in section \ref{sec:conclusions}.


\section{Relevant properties of the gyrokinetic equation}
\label{sec:GKE}

Here we review the electrostatic gyrokinetic turbulence model for a flux tube to identify the geometric features that appear.
A similar discussion can be found in \cite{jorge2020use}.
The magnetic field can be written $\mathbf{B}=\nabla\psi\times\nabla\alpha$ where $2\pi\psi$ is the toroidal flux and $\alpha$ is a field line label.
Coordinates $(x,y,z)$ are introduced where $z$ is the arclength along the field line, satisfying $\mathbf{B}\cdot\nabla z=B$, and $x=x(\psi)$ and $y=y(\alpha)$ are functions of $(\psi,\alpha)$ that each resemble a distance. 
(The results of this section are independent of the exact choice of these two functions.)
We consider a flux tube, for which the domain's small extent in $x$ and $y$ is on the order of a few gyroradii, much smaller than typical equilibrium scales like the minor radius.
Therefore $x$ and $y$ (or $\psi$ and $\alpha$) can be considered fixed for all equilibrium quantities.
However we do care about gyroradius-scale fluctuations in the distribution function and electrostatic potential, so the $x$- and $y$-dependence of these perturbations is retained.
Fluctuating quantities are taken to vary with $z$ on the same scale length as the equilibrium, and the extent of the flux tube domain in $z$ is of a comparable scale.

The fluctuating electrostatic potential $\Phi$ is Fourier expanded as
\begin{equation}
    \Phi(x,y,z,t) = \sum_{\mathbf{k}}\hat{\Phi}_{\mathbf{k}}(z,t) \exp(i k_x x + i k_y y),
\end{equation}
where $\mathbf{k}$ denotes $(k_x,k_y)$.
The distribution function of species $s$ is expanded as
\begin{equation}
    f_s = F_{Ms} - \frac{e_s \Phi}{T_s} F_{Ms} + h_s,
    \label{eq:h_def}
\end{equation}
where $F_{Ms} = n_s (2\pi)^{-3/2} v_s^{-3} \exp(-v^2 / (2 v_s^2))$ is the leading-order Maxwellian with density $n_s$ and temperature $T_s$, $v$ is the speed, $v_s = \sqrt{T_s / m_s}$ is the thermal speed, and $e_s$ is the species charge.
The nonadiabatic part of the distribution, $h_s$, has a Fourier expansion
\begin{equation}
    h_s(X,Y,z,v_{||},\mu,t) = \sum_{\mathbf{k}}\hat{h}_{s,\mathbf{k}}(z,v_{||},\mu,t) \exp(i k_x X + i k_y Y),
\end{equation}
where $(X,Y)$ are the values of $(x,y)$ at the guiding center position, $\mu=v_{\perp}^2/(2B)$, and $v_{||}$ and $v_{\perp}$ are the velocity components along $\mathbf{B}$ or perpendicular to it.
The nonadiabatic distribution is computed by evolving the gyrokinetic equation \citep{frieman1982nonlinear},
\begin{equation}
    \frac{\partial \hat{h}_{s,\mathbf{k}}}{\partial t}
    + v_{||} \frac{\partial \hat{h}_{s,\mathbf{k}}}{\partial z}
    -\mu \frac{\partial B}{\partial z} \frac{\partial \hat{h}_{s,\mathbf{k}}}{\partial v_{||}}
    + \mathbf{v}_d \cdot\nabla \hat{h}_{s,\mathbf{k}}
    +\mathcal{N}_{s,\mathbf{k}}
    = \frac{e_s J_{0,\mathbf{k}} F_{Ms}}{T_s} 
    \left( \frac{\partial \hat{\Phi}_{\mathbf{k}}}{\partial t} + i \omega_{*s}^T \hat{\Phi}_{\mathbf{k}} \right)
    + C_{s,\mathbf{k}}.
    \label{eq:GKE}
\end{equation}
Here, $\mathbf{b}=B^{-1} \mathbf{B}$ is the unit vector along the magnetic field, $\omega_{*s}^T=\omega_{*s} [1 + \eta_s (m_s v^2 / 2 T_s - 3/2)]$, and $\eta_s = d \ln T_s / d \ln n_s$.
Also, $\omega_{*s}= [\sigma k_y T_s / (e_s B_{ref})] d \ln n_s / d x$, and $\sigma = (B_{ref}/B^2)\mathbf{B}\cdot\nabla x\times\nabla y = B_{ref}(dx/d\psi)(dy/d\alpha)$ is constant over the flux tube domain to leading order. 
A constant reference field strength is denoted $B_{ref}$, and $C_{s,\mathbf{k}}$ is the gyroaveraged collision operator.
The factor $J_{0,\mathbf{k}}$ is shorthand for the Bessel function $J_0(k_\perp v_\perp / \Omega_s)$ where $k_\perp = | k_x \nabla x + k_y \nabla y|$, and $\Omega_s=e_s B / m_s$ is the gyrofrequency.
Also, the magnetic drift is
\begin{equation}
\label{eq:v_drift}
\mathbf{v}_d = \frac{m_s v_{\perp}^2}{2 e_s B^3} \mathbf{B}\times\nabla B
+ \frac{m_s v_{||}^2}{e_s B^2} \mathbf{B}\times\mathbf{\kappa},
\end{equation}
where $\mathbf{\kappa}=\mathbf{b}\cdot\nabla\mathbf{b}$ is the curvature, and $\mathcal{N}_{s,\mathbf{k}}$ denotes the nonlinear term:
\begin{equation}
    \mathcal{N}_{s,\mathbf{k}}=
    \sum_{\mathbf{k}'}
    \frac{\sigma J_{0,\mathbf{k'}} \hat{\Phi}_{\mathbf{k}'} \hat{h}_{s,\mathbf{k}''}}{B_{ref}}
    (k'_y k''_x - k'_x k''_y)
\end{equation}
where $\mathbf{k}''=\mathbf{k} - \mathbf{k}'$.
Note that where $v^2$ is needed in $F_{Ms}$ and $\omega_{*s}^T$, it can be computed from the independent variables via $v^2=v_{||}^2+2\mu B$, so the magnetic geometry enters via $B$.

The system is closed with the quasineutrality equation:
\begin{equation}
\hat{\Phi}_{\mathbf{k}}
\sum_s \frac{e_s^2 n_s }{T_s}
=    \sum_s e_s \int d^3v \; J_{0,\mathbf{k}}  \hat{h}_{s,\mathbf{k}} .
\end{equation}
The integral over velocity space for these velocity coordinates $\int d^3v = 2\pi \int_{-\infty}^{\infty} dv_{||} \int_0^{\infty} d\mu \,B$ depends on the magnetic field via $B$.

Once the potential and distribution function are computed, we are principally interested in the turbulent heat flux for the ions, $s=i$:
\begin{equation}
    Q = \sum_{\mathbf{k}}
    \left\langle \int d^3v \frac{m_i v^2}{2}
    \hat{h}_{i,\mathbf{k}} \frac{i k_y \sigma J_{0,\mathbf{k}} \hat{\Phi}_{-\mathbf{k}}}{B_{ref}}
    \right\rangle,
\end{equation}
where $\left \langle\ldots\right\rangle$ denotes a flux surface average: $\langle u \rangle = \left(\int d\ell/B\right)^{-1} \int d\ell\, u/B$ for any quantity $u$.

Now, let us examine the places where the magnetic geometry enters the above model.
From the $\mathbf{v}_d$ term in (\ref{eq:GKE}), which can be expanded as
\begin{equation}
    \left(\mathbf{v}_d \cdot\nabla x \right) i k_x \hat{h}_{s,\mathbf{k}}
    + \left(\mathbf{v}_d \cdot\nabla y \right) i k_y \hat{h}_{s,\mathbf{k}},
\end{equation}
we see that the $\nabla x$ and $\nabla y$ components of the magnetic drift (\ref{eq:v_drift}) appear.
Note that $\mathbf{b}\times\nabla B\cdot\nabla\psi = \mathbf{B}\times\mathbf{\kappa}\cdot\nabla\psi$ exactly for an MHD equilibrium, even if the ratio $\beta$ of thermal to magnetic pressure is not small, so the two terms in $\mathbf{v}_d\cdot\nabla x$ can be combined as a multiple of $\mathbf{b}\times\nabla B\cdot\nabla x$.
Next, in the argument of the Bessel functions,
\begin{equation}
    k_\perp = \sqrt{k_x^2 |\nabla x|^2 + 2k_x k_y \nabla x\cdot\nabla y + k_y^2 |\nabla y|^2},
\end{equation}
the quantities $|\nabla x|^2$, $\nabla x\cdot\nabla y$, and $|\nabla y|^2$ appear.
Finally, $B$ appears in numerous places: through $\Omega_s$ in the Bessel functions, in the flux surface average in the heat flux, in $v(v_{||}, \mu B)$, in $\int d^3v$, and via $\partial B/\partial z$ in (\ref{eq:GKE}).
Thus, we find that the raw geometric features entering the gyrokinetic model are the following seven functions of $z$:
\begin{equation}
    B, 
    \;
    \frac{\mathbf{B}\times\nabla B\cdot\nabla x}{B^3},
    \;
    \frac{\mathbf{B}\times\nabla B\cdot\nabla y}{B^3},
    \;
    \frac{\mathbf{B}\times \mathbf{\kappa}\cdot\nabla y}{B^2},
    \;
    |\nabla x|^2,
    \;
    \nabla x \cdot\nabla y,
    \;
    |\nabla y|^2.
    \label{eq:raw_features}
\end{equation}
As stated previously, the $x$ and $y$ variation of these equilibrium quantities over the flux tube simulation domain is negligible due to the small extent of the domain in these coordinates, so we only need the variation of these quantities in $z$ along a field line.
If a parallel coordinate $z$ other than arclength is used, there would be an additional geometric input $\mathbf{b}\cdot\nabla z$ required in the parallel streaming term of (\ref{eq:GKE}).

\begin{figure}
  \centering
   \includegraphics[width=2.5in]{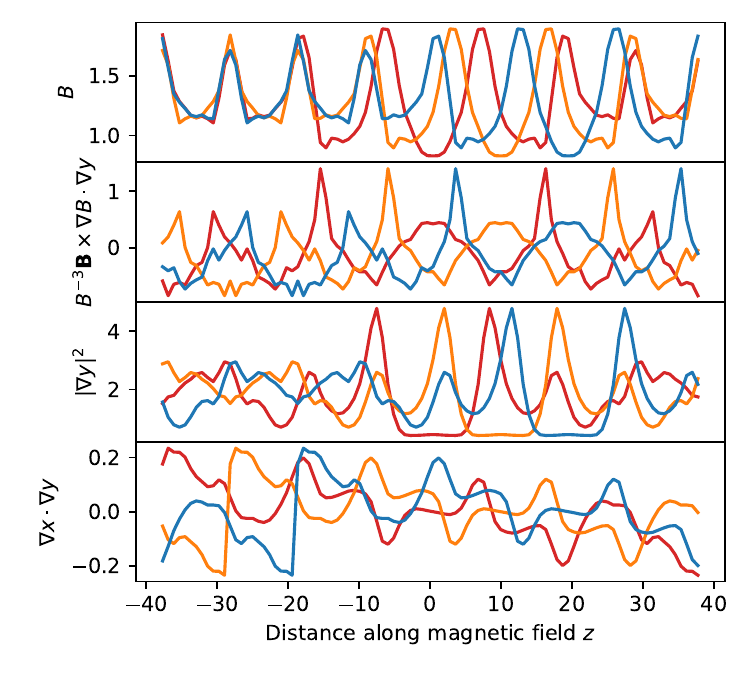}
   \includegraphics[width=2.5in]{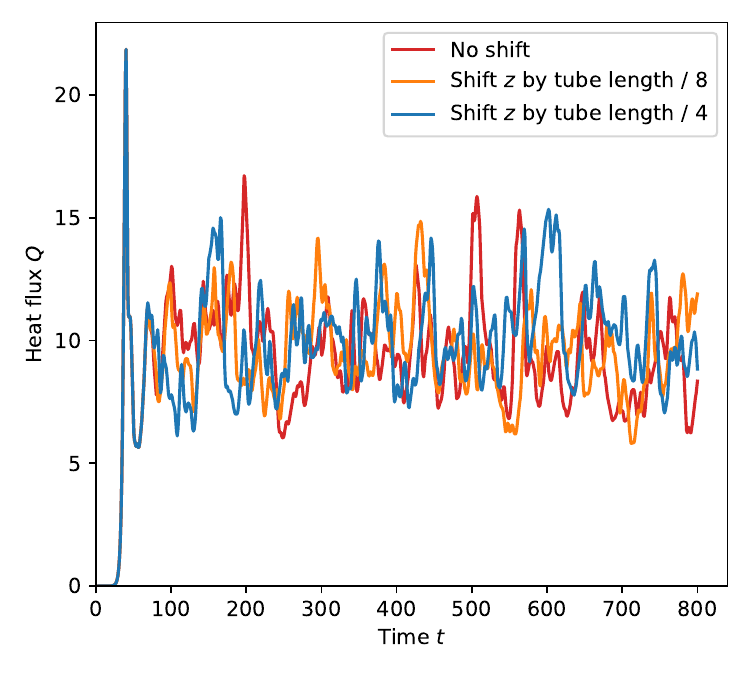}
  \caption{Translation-invariance of the gyrokinetic-quasineutrality system.
  Left: A periodic translation in $z$ is applied to all of the $z$-dependent inputs to the gyrokinetic system, eq (\ref{eq:raw_features}).
  Only four of the seven are shown for simplicity, but all are translated.
  Right: The average heat flux is unchanged by the translation.
  }
\label{fig:translation_invariance}
\end{figure}

We can now understand a translation-invariance property of the gyrokinetic model above, which ML models should preserve.
The key idea is that $z$ does not appear explicitly anywhere in the model - $z$-dependence in the equations enters only through the functions in (\ref{eq:raw_features}).
To precisely state the translation-invariance property, suppose the gyrokinetic-quasineutrality system is solved with periodic boundary conditions in $z$.
Then the heat flux is exactly unchanged if a periodic shift $f(z) \to f(z + \Delta)$ is applied where $f$ indicates each of the seven raw features (\ref{eq:raw_features}) along with $\Phi$ and $h_i$, where the shift $\Delta$ is the same for all quantities.
If the seven raw features are shifted in this way but the initial conditions for $\Phi$ and $h_i$ are not, the detailed dynamics will change, but under the usual assumption that the mean heat flux is independent of the initial condition, the mean heat flux would still be invariant.
We confirmed that this translation-invariance indeed holds for nonlinear GX simulations, as shown in figure \ref{fig:translation_invariance}.
If a twist-and-shift boundary condition \citep{beer1995field, martin2018parallel} is used in $z$ instead of periodic boundary conditions, and a finite number of $k_x$ and $k_y$ modes are included, the translation-invariance will be slightly violated, because the outgoing distribution function is set to zero at the unlinked flux tube ends.
However if the simulation is well resolved with respect to the number of $k_x$ modes, the heat flux is evidently approximately independent of the number of times a tube segment links to itself, and hence approximately independent of where the outgoing distribution function is set to zero.
Thus the breaking of translation-invariance should be small.

For all results that follow, the raw features (\ref{eq:raw_features}) are normalized by a reference length $L_{ref}$ and reference field strength $B_{ref}$.
The effective minor radius $a$ is adopted for $L_{ref}$, and $B_{ref}$ is defined by equating $\pi a^2 B_{ref} = 2\pi|\psi_a|$, where $\psi_a$ is the value of $\psi$ at the equilibrium boundary.
We choose the perpendicular coordinates to be $x=a\sqrt{\psi/\psi_a}$ and $y=-\alpha x \, \mathrm{sign}(\psi_a)$, so $\sigma=-1$.


\section{Dataset generation}
\label{sec:data_generation}

In this section, we first present the method used to generate stellarator MHD equilibria with diverse geometries.
Next, details of the nonlinear gyrokinetic simulations in these equilibria are given.
After describing some general properties of the dataset, a few configurations are highlighted that are interesting due to their extreme values of the heat flux.


\subsection{Magnetohydrodynamic equilibria}
\label{sec:equilibrium_generation}

An MHD equilibrium is determined by the boundary shape together with two functions of the toroidal flux, typically the pressure and enclosed toroidal current \citep{kruskal1958equilibrium}.
It is not obvious how to best sample this space, particularly the space of boundary shapes.
Boundaries with self-intersections or other such pathologies should be avoided.
Also, a compromise must be struck between preserving similarity of the shapes to ``real'' devices (built experiments or theoretical configurations designed through serious optimization) while also allowing for new possible geometries.

\begin{figure}
  \centering
   \includegraphics[width=\columnwidth]{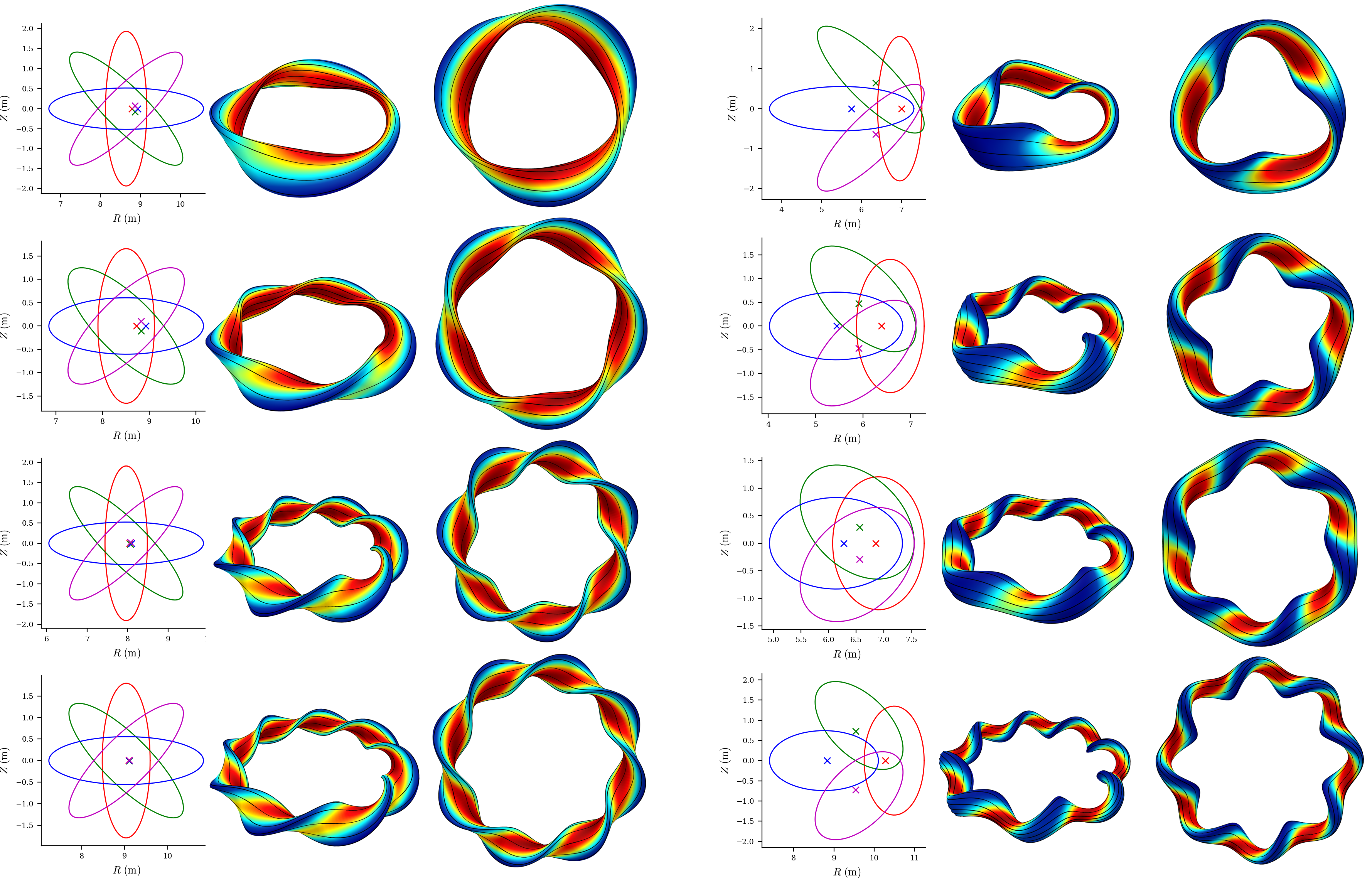}
  \caption{Examples of the rotating-ellipse equilibria included in the dataset.
  2D plots show the cross sections at which the toroidal angle is 0, $1/4$, $1/2$, and $3/4$ of a field period.
  3D images show each configuration from two angles\changed{, with color indicating $|B|$ (red = high, blue = low) and field lines in black}.
  Left columns: configurations in which the boundaries are centered on a circle.
  Right columns: configurations in which the boundaries are centered on a curve with torsion.
  }
\label{fig:heliotron_equilibria}
\end{figure}

To balance these considerations, we assemble a set of 23,577 equilibria drawn from three groups.
The first group has heliotron-like rotating ellipse shapes, in which parameters of the shape (number of field periods, aspect ratio, elongation, axis torsion, beta) are sampled randomly.
Equilibria in the second group are taken from the QUASR database of quasi-axisymmetric (QA) and quasi-helically symmetric (QH) configurations \citep{giuliani2024direct, giuliani2024comprehensive}.
We use both the original QUASR configurations, which are all vacuum fields, and also generate new configurations by adding pressure while keeping the plasma boundary shape fixed.
For the third group, random boundary shapes are generated by sampling Fourier modes that have been fit to a dataset of previous stellarator shapes.
The combined set of configurations includes values of aspect ratio ranging from 2.9 to 10, volume-averaged beta from 0 to 5\%, and number of field periods from 2 through 8.
Thus, the set of equilibria is diverse, and includes both omnigenous and non-omnigenous geometries.
All equilibria have the same minor radius and same boundary toroidal flux, resulting in the same normalizing length and normalizing field strength, so the gyro-Bohm normalizations are identical.
Examples of equilibria from the three classes are shown in figures \ref{fig:heliotron_equilibria}-\ref{fig:random_equilibria}.
More details of the procedures for generating equilibria are given in appendix \ref{app:equilibrium_details}.

\begin{figure}
  \centering
   \includegraphics[width=\columnwidth]{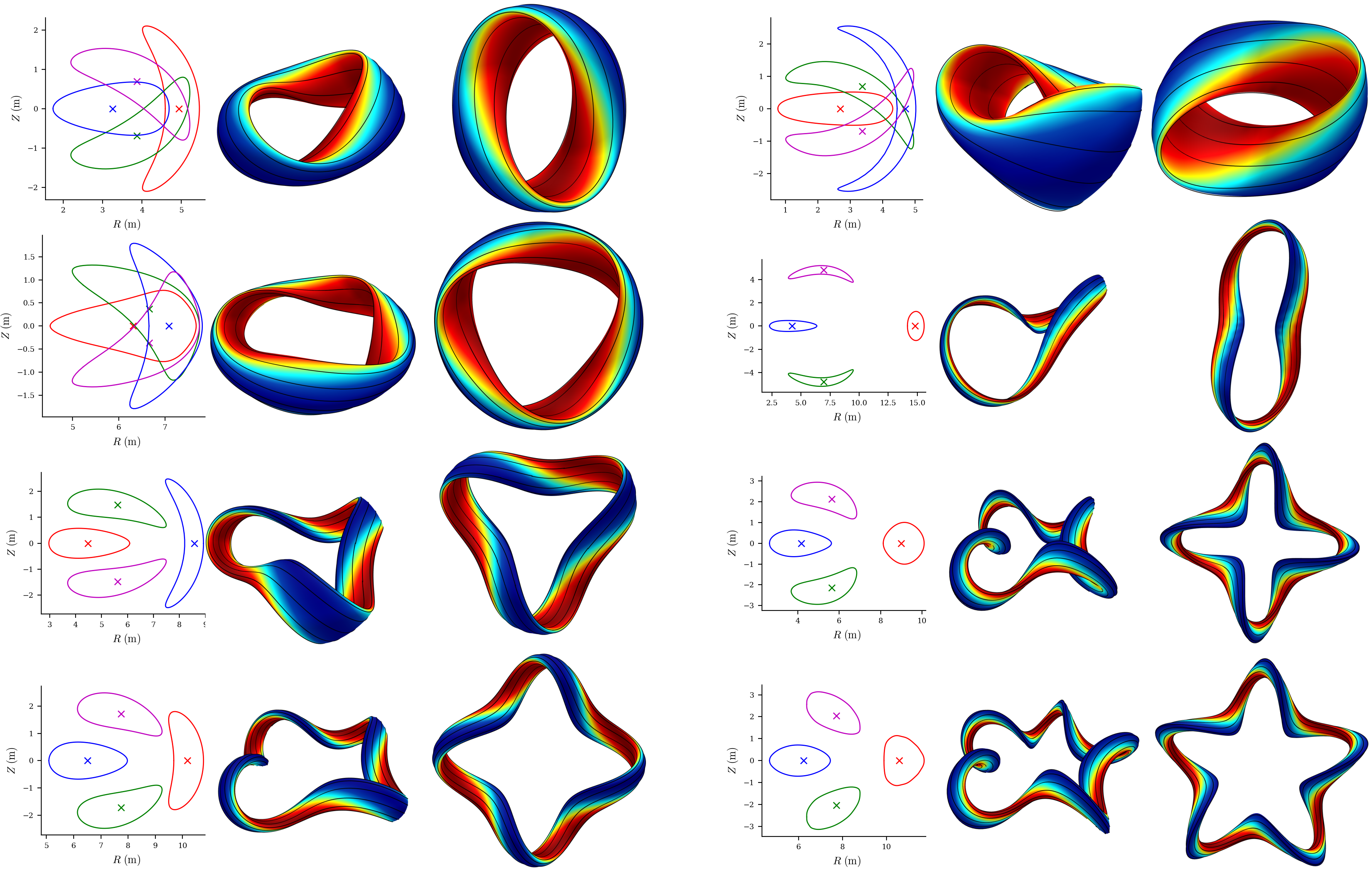}
  \caption{Examples of the QUASR quasi-axisymmetric and quasi-helically symmetric equilibria included in the dataset.
  2D plots show the cross sections at which the toroidal angle is 0, $1/4$, $1/2$, and $3/4$ of a field period.
  3D images show each configuration from two angles\changed{, with color indicating $|B|$ (red = high, blue = low) and field lines in black}.
  }
\label{fig:QUASR_equilibria}
\end{figure}

\begin{figure}
  \centering
   \includegraphics[width=\columnwidth]{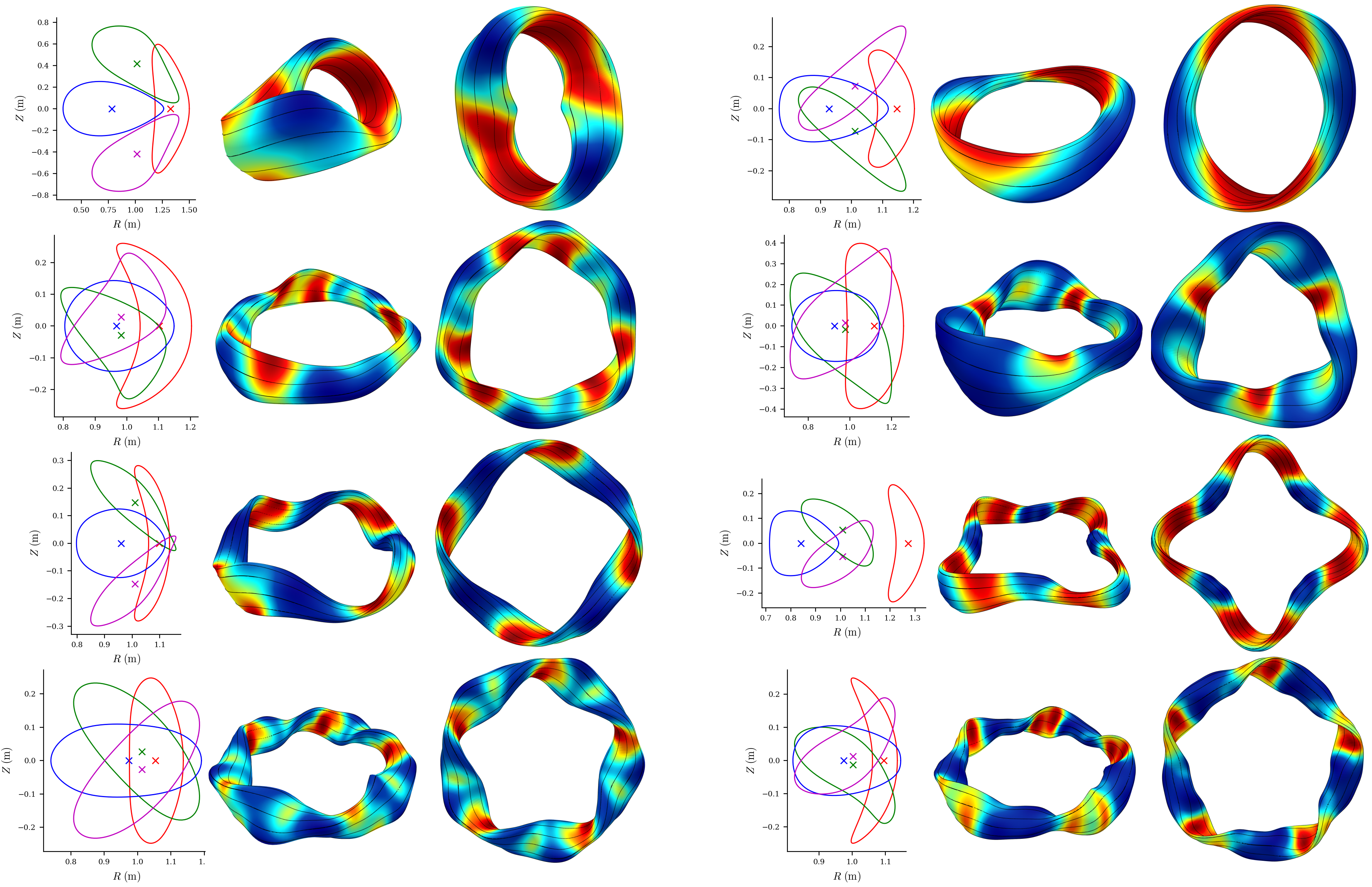}
  \caption{Examples of the equilibria generated with random boundary Fourier modes.
  2D plots show the cross sections at which the toroidal angle is 0, $1/4$, $1/2$, and $3/4$ of a field period.
  3D images show each configuration from two angles\changed{, with color indicating $|B|$ (red = high, blue = low) and field lines in black}.
  }
\label{fig:random_equilibria}
\end{figure}


\subsection{Gyrokinetic simulations}

In each equilibrium, we extract at least four flux tubes, more in the QUASR vacuum configurations so as to include more omnigenous geometries.
The radial location $\rho=\sqrt{s}$ for each flux tube is sampled randomly from $[0, 1]$, where $s=\psi/\psi_a$ is the normalized toroidal flux.
All flux tubes are stellarator-symmetric, with at least one flux tube in every configuration centered on each of the points $(\theta,\phi)=(0,0)$, $(\pi,0)$, $(0,\pi/\nfp)$, and $(\pi,\pi/\nfp)$.
Nothing in the procedure limits the method to stellarator-symmetric equilibria or flux tubes.
The total number of flux tube geometries in the dataset is 100,705.
The length of the flux tubes in real space was chosen to be the same for every configuration, 75 times the minor radius, which for aspect ratio 6 corresponds to roughly two full toroidal transits.
Due to the variation in $\iota$, the number of poloidal transits varied between configurations.
By making all flux tubes have the same physical length, distances in the $z$ coordinate can be compared meaningfully between flux tubes, and for each integer $j$, the $j$th Fourier series coefficients in all flux tubes correspond to the same physical $k_{||}$.
These properties may be advantageous for extracting patterns in the dataset.

Nonlinear electrostatic simulations with adiabatic electrons were then carried out with the GX code \citep{mandell2024gx}.
We run GX twice for each flux tube, resulting in two datasets each of size $N=100,705$.
In the first dataset, the temperature and density gradients were fixed to the same values for all flux tubes.
In the second dataset, these gradients were chosen randomly for each flux tube.
Using the latter dataset, models can potentially find interactions between geometry and gradients in their effect on the heat flux; for example, the dependence on geometry may differ close to the critical gradient compared to far above the critical gradient.
However we also assembled the dataset with fixed gradients since it enables several of the analyses in section \ref{sec:features} and allows us to focus cleanly on the effects of geometry.
For the fixed-gradient dataset, we take $a/L_{Ti} = 3$ and $a/L_n=0.9$, \changed{where $a/L_{Ti} = -(a/T_i) dT_i/dx$ and $a/L_n=-(a/n) dn/dx$ are the normalized gradients of ion temperature and density,} reflecting typical measurements at the $s=0.5$ surface of W7-X \citep{beurskens2021ion,lunsford2021characterization, zhang2023observation}.
For the varying-gradient dataset, each simulation has $a/L_{Ti}$ and $a/L_n$ sampled randomly within plausible experimental ranges.

Additional details of the turbulence simulations are given in appendix \ref{app:turbulence_details}.


\subsection{Configurations with very low or high heat flux}

\begin{figure}
  \centering
   \includegraphics[width=3.5in]{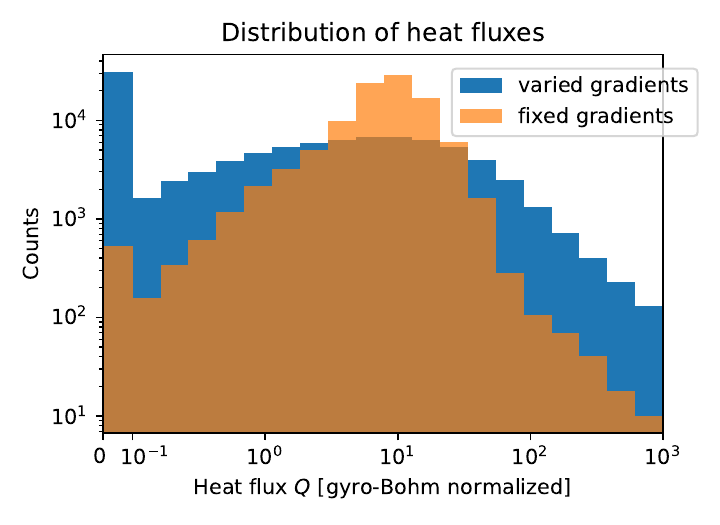}
  \caption{Distribution of heat fluxes for the fixed-gradient and varied-gradient datasets.
  In the latter, 30\% of the simulations were stable, with $Q \approx 0$.
  Simulations with $Q<0.1$, considered stable, are included in the leftmost bar.
  }
\label{fig:Q_distribution}
\end{figure}

The distribution of heat fluxes for the dataset is shown in figure \ref{fig:Q_distribution}.
It is striking that even for the dataset with fixed gradients, $Q$ varies by over four orders of magnitude over the data.
This variation is evidently due purely to the geometry.

\begin{figure}
  \centering
   \includegraphics[width=\columnwidth]{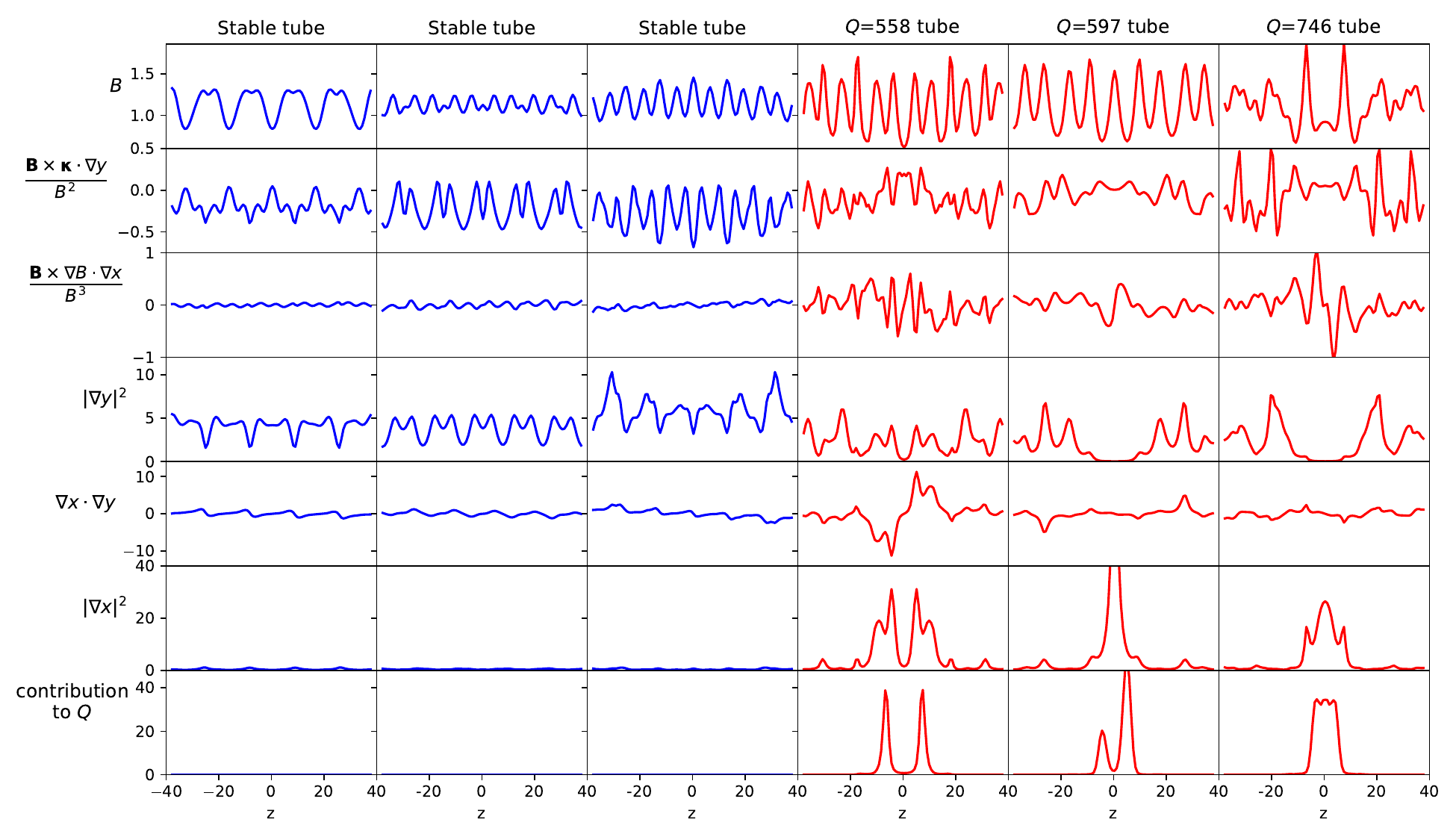}
  \caption{
  Some trends are apparent between flux tubes with very low or high heat flux.
  The columns show six flux tubes from the $n_{fp}=3$ equilibria with random boundary Fourier modes, the first three stable, the last three with very high $Q$ at the same gradients.
  The top six rows are the inputs to the gyrokinetic-quasineutrality system ($B^{-3}\mathbf{B}\times\nabla B\cdot\nabla y$ is omitted for simplicity since it is similar to $B^{-2}\mathbf{B}\times\mathbf{\kappa}\cdot\nabla y$), while the bottom row shows the contribution to the total heat flux vs $z$.
  }
\label{fig:low_high_flux_configs}
\end{figure}

It is interesting to examine configurations with particularly low or particularly high heat flux.
Several such flux tubes are displayed in figure \ref{fig:low_high_flux_configs}.
The columns represent six flux tubes from the \changed{dataset. In}  the fixed gradient dataset with $a/L_{T}=3$ and $a/L_n=0.9$, the first three tubes are stable, while the other three have $Q > 500$, even though the gradients are identical.
These flux tubes are all taken from $n_{fp}=3$ equilibria from the group with random boundary Fourier modes.
The first six of the seven rows show the geometric inputs to the gyrokinetic-quasineutrality system.
To simplify the figure, $B^{-3}\mathbf{B}\times\nabla B\cdot\nabla y$ is not shown since it is similar to $B^{-2}\mathbf{B}\times\mathbf{\kappa}\cdot\nabla y$.
The bottom row shows the contribution to the heat flux $Q$ as a function of $z$.
For each of the seven rows, the vertical scales are the same for each column to allow comparison.

Several patterns are apparent.
The configurations with highest heat flux have regions with very large $|\nabla x|^2$.
The high-heat-flux tubes also have larger magnitudes of $B^{-3}\mathbf{B}\times\nabla B\cdot\nabla x$.
The stable tubes have more negative values of $B^{-2}\mathbf{B}\times\mathbf{\kappa}\cdot\nabla y$, meaning mostly good curvature.
All of these patterns foreshadow findings in section \ref{sec:features}.

\changed{
For the geometry-gradient pairings with very large heat flux, where the gyroBohm-normalized $Q$ is not small compared to $1 / {\rho_*}$, with $\rho_*$ the ratio of typical gyroradius to macroscopic scale, the ordering used to derive the gyrokinetic model may no longer be accurate.
Therefore these simulation runs may not accurately describe real physical scenarios.
Nonetheless it is still reasonable to consider the $\delta f$-gyrokinetic model as an abstract mathematical function, a map from seven functions (the geometry) and two real numbers (the gradients) to a real number ($Q$), and to develop approximations of this function.
Moreover, since surrogates will be more accurate at interpolation than extrapolation, it is desirable to include at least a few of these extreme data points in the dataset.
}


\subsection{Weighting of errors for regression}

When fitting regression models to the heat flux data and assessing their accuracy, some care is required to measure and weigh errors appropriately.
The first issue is that the heat flux can span several orders of magnitude.
If a model makes an absolute prediction error of 1 in gyro-Bohm units, this error is significant if the true heat flux is 0.5, but it is a relatively minor error if the true heat flux is 100.
Moreover, the standard deviation (over time) of the heat flux within one simulation tends to scale approximately linearly with the mean heat flux, so high-$Q$ cases have higher uncertainty than low-$Q$ cases.
For these reasons, it is natural to measure errors in a relative rather than absolute sense.
Weighing errors this relative sense can be achieved by performing regression on $\ln Q$ rather than $Q$ directly.
In particular, models should be scored based on the mean squared error or coefficient of determination $R^2$ evaluated using $\ln Q$.
\changed{Thus, in the standard definition $R^2 = 1-\left[\sum_j (y_j-f_j)^2\right]/\left[\sum_j (y_j - \bar{y})^2\right]$ where $y_j$ are the true target values, $\bar{y}$ is their mean, and $f_j$ are the predicted values, one would take $y_j$ and $f_j$ to be the true and predicted values of $\ln Q$ rather than of $Q$ itself.}

However, an additional important consideration is that some simulations are stable.
In a stable simulation, the heat flux is typically slightly positive but $\ll 1$ at the end of the computation due to decay of the initial perturbation, and the exact value of $Q$ is not very meaningful.
There is not a steady state to average over, and the final $Q$ is influenced by the magnitude of the initial condition.
If the heat flux from a simulation is $Q=10^{-12}$ and a regression model predicts it to be $10^{-4}$, the model is accurately predicting that the system is stable (or nearly so), but scoring the model based on the error in $\ln Q$ would treat the prediction error to be large.
In some stable simulations where $|Q|$ has very small magnitude, $Q$ may even be slightly negative, in which case $\ln Q$ cannot even be evaluated.

There are several possible approaches to account for these considerations.
One approach is to consider a classification problem for stability vs instability separately from the regression problem.
In this case, the regression only needs to be evaluated if the classifier has first predicted instability, i.e. in the unstable region of parameter space, so the regression can be fit using only the subset of the data in which $Q$ exceeds some small threshold.
Here we adopt this approach with a threshold of 0.1, and our experience is that results are insensitive to the exact choice of threshold.
With this approach, the training set does not include any points with $Q < 0.1$ or negative $Q$, so it is straightforward to use $\ln Q$ as the regression target.
The full dataset is still used for training the classifier.
If the classifier is interpretable, it directly provides insight into the factors that determine the critical gradient.

Several other approaches are possible.
As in \cite{van2020fast}, a regression could be fit by minimizing a custom objective that penalizes errors in $Q$ (or $\ln Q$) only in the unstable region, while in the stable region, only penalizing positive values of $Q$, to prevent spurious prediction of instability.
However, not all regression methods allow for a custom objective, e.g. nearest-neighbors.
Or, regression could be performed on the full dataset with a standard objective after all values of $Q$ below a threshold are replaced by the threshold.
While straightforward to implement, this method introduces non-smoothness in the predicted $Q$; this is not a problem for decision tree models but it may introduce inaccuracies with smooth models like neural networks.
A fourth option is to perform regression on the full dataset using the target quantity
\begin{equation}
\hat{Q} = \ln(1 + Q).
\label{eq:ln_1_plus_Q}
\end{equation}
For $Q \gg 1$, the quantity $\hat{Q}$ behaves like $\ln Q$, while for $Q \ll 1$, $\hat{Q}$ varies approximately linearly with $Q$.
Therefore a regression model that is fit by minimizing the mean squared error in $\hat{Q}$, or $R^2$ computed from $\hat{Q}$, prediction errors will effectively be based on relative error in $Q$ for large $Q$, but based on absolute error for small $Q$, as desired.
We have tried the first, third, and fourth of these methods, and our experience is that they give very similar results.
For the rest of the discussion we use the first method.


\section{Convolutional neural network models}
\label{sec:CNN}

We develop neural network surrogates to predict heat flux values from the simulations described above. Our approach involves two key strategies: (i) designing a neural network that remains invariant to cyclic permutations and (ii) ensuring reliable predictions through an ensemble of the top-k models selected via hyperparameter optimization.

\subsection{Cyclic invariant neural network}
Our goal is to develop a surrogate model based on a neural network that maintains invariance to cyclic permutations inherent to the gyrokinetic system. 
An analogy exists with computer vision research, where translation invariance is important as well: if a cat is present in an image, it should be recognized as a cat even if its location within a bitmap is shifted.
A standard approach to achieving approximate translation-invariance in computer vision is through convolutional (as opposed to fully-connected) neural networks.
This type of neural network contains convolutional layers, in which the spatial data are convolved with a kernel, which is a translation-equivariant operation.
Between convolutional layers, pooling layers then reduce the spatial resolution, typically by a factor of two, and a sufficient number of alternating convolutional and pooling layers converts the spatial data to approximately translation-invariant features.
While computer vision applications use two-dimensional convolution and pooling operations, for the turbulence application here we will use one-dimensional convolution and pooling.

During training, we further reinforce the translation-invariance by augmenting the dataset with a sufficient number of sequences that are randomly permuted (cyclically) in $z$, allowing the network to learn patterns that are independent of any particular alignment with respect to $z$. This strategy prevents the model from developing biases toward specific alignments, ultimately improving generalization and stability.

Another key aspect of our surrogate model development is the use of ensemble learning, a powerful machine-learning technique that aggregates predictions from multiple base models to improve overall performance. Ensemble methods have proven to be highly effective in reducing prediction variance, mitigating model biases, and improving generalization, making them well-suited for applications involving complex, high-dimensional data such as ours.
By leveraging an ensemble of multiple models, we aim to construct a predictive framework that is not only more accurate but also more stable when encountering unseen data. The core advantage of ensemble learning is its ability to balance trade-offs between different models—some models may be more sensitive to specific features, while others may capture different patterns. By combining them, we create a more comprehensive representation of the underlying physical system.

To implement this strategy, we design our base model to be highly flexible, allowing variations in neuron count, layer depth, kernel shape, activation functions, etc., also known as hyperparameters. This flexibility is essential for tuning the model to achieve optimal predictive performance across different configurations. However, selecting the best architecture requires careful exploration of the hyperparameter space.
To efficiently search for the best hyperparameters, we employ DeepHyper \citep{balaprakash2018deephyper, egele2023asynchronous}, an advanced hyperparameter optimization tool designed for deep learning. DeepHyper allows us to automate the search for high-performing  configurations, reducing manual tuning efforts and accelerating model development. We will discuss the details of this optimization process in subsequent sections.

\begin{figure}
  \centering
  \includegraphics[width=0.8\columnwidth]{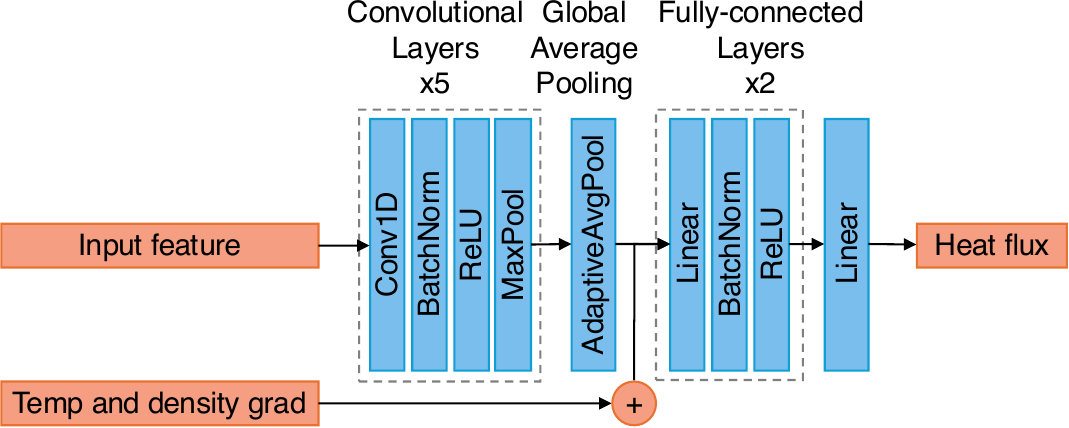}
  \caption{The surrogate model architecture to learn heat flux averages using a structured neural network. It consists of three main components: i) Feature Extraction with Conv1D, ii) Global Average Pooling, and iii) Fully Connected Layers.}
\label{fig:model}
\end{figure}

With these considerations in mind, we design our surrogate neural network using a structured approach that consists of three main components, shown in Figure \ref{fig:model}:
\begin{enumerate}
    \item {\bf Feature Extraction with Convolutional Layers}: The first component comprises $n$ consecutive layers of neural network blocks, each containing 1D convolutional layers (Conv1D). These layers are responsible for extracting spatial features from the input sequences while preserving cyclic invariance. Each convolutional layer processes $x$ data channels and outputs $y$ number of transformed channels, which are subsequently normalized using batch normalization to stabilize training and improve convergence. A max-pooling layer follows each block to downsample the extracted features, reducing computational complexity while retaining the most relevant information.
    \item {\bf Global Average Pooling (GAP) for Dimensionality Reduction}: To further condense the extracted feature maps, we integrate global average pooling (GAP). Unlike traditional pooling methods that operate on small local regions, GAP computes the average activation value for each feature map across all spatial dimensions. This not only reduces the number of parameters but also ensures that the model focuses on high-level feature representations rather than localized details, making it particularly effective for cyclic data.
    \item {\bf Fully Connected Layers for Prediction}: The final stage consists of $m$ fully connected linear layers that process the compact feature representation and map it to the target variable: predicted heat flux averages. These layers refine the extracted features and produce the final output by capturing complex, non-linear relationships in the data.
    Since the temperature and density gradient are not functions of $z$, these two inputs bypass the convolutional layers and feed directly to the first fully connected layer.
\end{enumerate}

\subsection{Hyper parameter optimizations and ensemble learning}

\begin{table}
  \begin{center}
\def~{\hphantom{0}}
  \begin{tabular}{rrrl}
    Hyperparameter & Count  & Range & Description \\
    \verb|kernel_size| & 5 & (2, 16) & Conv1D's convolving kernel size \\
    \verb|conv_channel| & 5 & (4, 32) & Conv1D's number of output channels \\
    \verb|fc_dimension|  & 2 & (4, 32) & Fully-connected layer's output feature size \\
    \verb|batch_size|   & 1 & (16, 64) & Batch size \\
    \verb|learning_rate|  & 1 & (1e-5, 1e-1) & Learning rate (sampled in log-uniform) \\
    \verb|scheduler_patience| & 1 & (5, 20) & ReduceLROnPlateau scheduler's patience parameter \\
    \verb|scheduler_factor|   & 1 & (0.1, 1.0) & ReduceLROnPlateau scheduler's factor parameter \\
  \end{tabular}
  \caption{List of hyperparameters and their search ranges explored using DeepHyper.}
  \label{tab:hplist}
  \end{center}
\end{table}

To enhance the predictive performance of our surrogate model, we employ ensemble learning, which aggregates predictions from multiple base models. By combining different models, we create a more robust and comprehensive representation of the underlying physical system, improving accuracy and generalization.

In our model, we carefully parameterize several key hyperparameters that influence performance. These include the kernel size of the 1D convolution layers, the number of output channels in Conv1D layers, the number of neurons in fully connected layers, batch size, learning rate, and scheduler patience for dynamically adjusting the learning rate. A summary of these hyperparameters is provided in Table \ref{tab:hplist}. We fix the number of Conv1D layers at five, so the associated pooling layers reduce the number of spatial points from 96 to $96/2^5=3$, and include two fully connected layers. While these constraints simplify the search space, the total number of possible hyperparameter configurations remains extremely large, exceeding one hundred quadrillion ($10^{17}$). Exploring this vast space through brute-force methods is computationally infeasible.

To efficiently navigate this large hyperparameter space, we leverage DeepHyper \citep{balaprakash2018deephyper, egele2023asynchronous}, an advanced hyperparameter optimization tool. DeepHyper systematically finds the best set of hyperparameters that maximize model accuracy using different search strategies including Bayesian Optimization, Evolutionary Strategies, and Random Search. Bayesian optimization uses a probabilistic approach to refining hyperparameters, while evolutionary strategies optimize parameters through genetic algorithms. Random Search serves as a baseline method. DeepHyper is designed for parallel execution in distributed computing environments such as clusters and supercomputers, making it effective for  hyperparameter optimization tasks at scale. 
In DeepHyper, Bayesian optimization (BO) that we adopt  follows a single-manager, multiple-worker scheme to explore and exploit the search space. The manager maintains a probabilistic surrogate model (e.g., Random Forest) to predict the performance of configurations and select promising candidates based on an acquisition function (e.g., Expected Improvement or Upper Confidence Bound). The workers asynchronously evaluate these candidates in parallel, returning results to update the surrogate model iteratively. To improve robustness and generalization, DeepHyper generates an ensemble of models by aggregating the top-performing configurations identified during the search.

In this study, we employ DeepHyper to explore the hyperparameter space of our base model, aiming to identify and select the top-N best models for ensemble learning. DeepHyper systematically searches the hyperparameter space to optimize model performance. To guide this optimization, we use the coefficient of determination $R^2$, as the objective function, ensuring the selected models achieve the highest predictive accuracy.
The base neural network model is implemented using pytorch.

\begin{figure}
  \centering
  \includegraphics[width=\columnwidth]{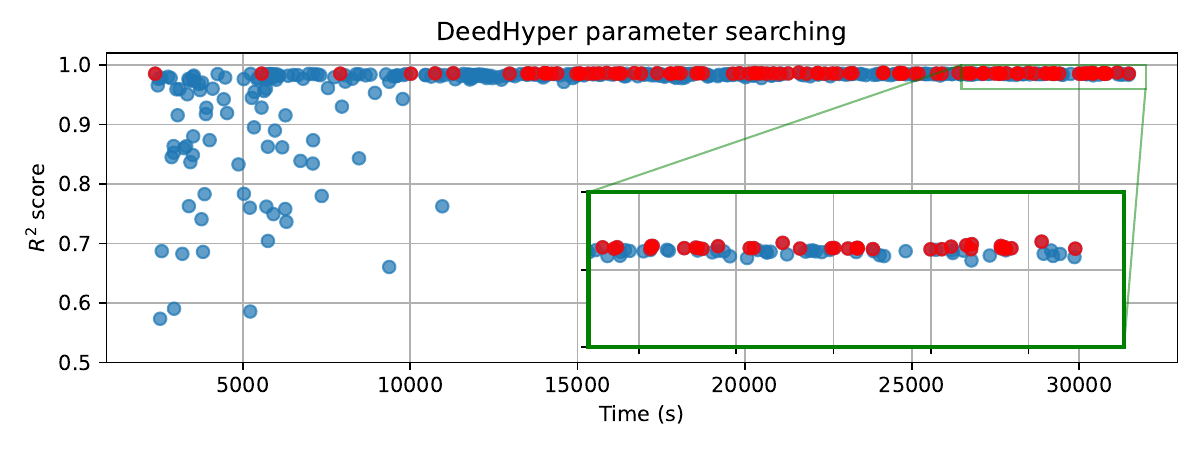}
  \includegraphics[width=\columnwidth]{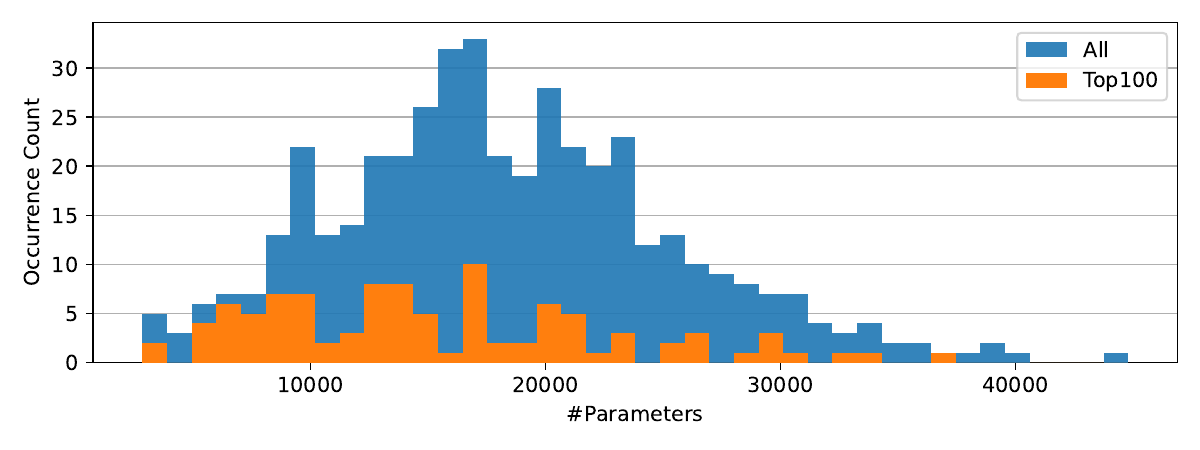}
  \caption{Results of the DeepHyper hyperparameter search. Each point represents a model with a unique hyperparameter configuration explored by DeepHyper, plotted against its completion time (x-axis) and performance score (y-axis). The search was conducted using 64 GPUs over approximately nine hours, evaluating a total of 443 models. The top 100 highest-performing models, selected for the final ensemble, are highlighted in red.
  The bottom plot shows the histogram of model sizes (number of parameters on the x-axis) for all 443 models explored by DeepHyper and the top 100 selected models.}
\label{fig:hpsearch}
\end{figure}

Figure \ref{fig:hpsearch} presents the results of our hyperparameter search. Each point in the figure represents a model explored during the optimization process. The x-axis denotes the time of completion, while the y-axis shows the model’s performance measured by the $R^2$ score. Over a period of approximately nine hours, we utilized 64 GPUs to conduct this search. DeepHyper evaluated 443 different models with varying hyperparameter configurations, systematically assessing their performance. After completing the search, we selected the top 100 models with the highest $R^2$ scores to form an ensemble, ensuring improved predictive accuracy and generalization.
The bottom plot presents a histogram of model sizes, represented by the number of parameters on the x-axis, for all 443 models explored by DeepHyper and the top 100 high-performing selected models, illustrating the distribution of model sizes before and after selection.

\begin{figure}
  \centering
  \includegraphics[width=0.64\columnwidth]{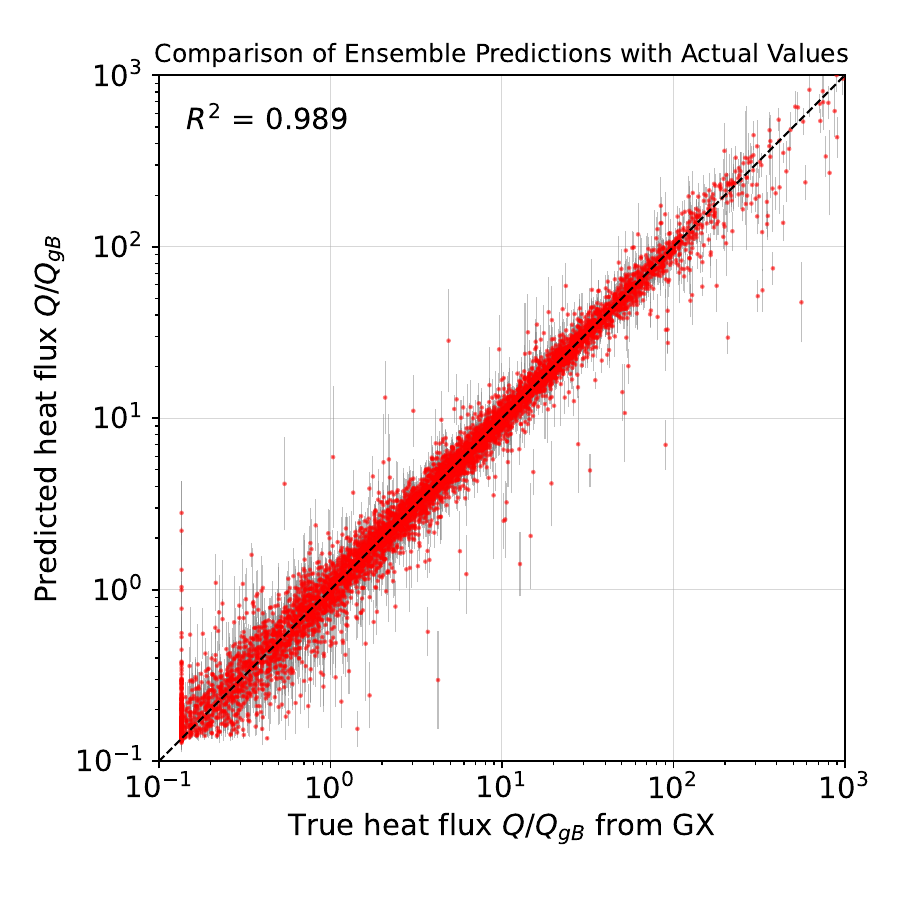}
  \caption{Prediction performance of an ensemble of the top 100 models selected from DeepHyper for the varied-gradient dataset. After exploring 443 models by DeepHyper, the top 100 were chosen based on their performance and evaluated against a test dataset of 9,785 samples. Each dot represents the mean prediction of the ensemble, while the vertical bars indicate ±1 standard deviation. The ensemble achieved an overall  $R^2$ score of 0.989, demonstrating strong predictive accuracy and stability.}
\label{fig:cnn}
\end{figure}

Figure \ref{fig:cnn} illustrates the prediction performance of this ensemble, which consists of the 100 highest-performing models from the DeepHyper optimization, on the varied-gradient data. Each red dot in the figure represents the mean prediction of the ensemble at each target value in the test dataset, while the vertical bars indicate the range of $\pm1\sigma$ (standard deviation) of predictions across the 100 models. We evaluated the ensemble using a total of 9,785 held-out test samples, achieving an overall $R^2$ score of 0.989, computed by treating the ensemble mean as the prediction.
Additionally, the neural networks exhibited efficient inference speed, taking a total 1,031.0 seconds on a single NVIDIA A100 GPU to process 9,785 test samples across the 100 models. 
To make a single prediction of the heat flux, this time corresponds to an average of 0.001 seconds for a single CNN and 0.1 seconds for the ensemble.
These times represent a factor of $\sim 4,000$ speed-up for evaluating the ensemble compared to one of the gyrokinetic simulations used to generate the training data; for evaluating a single CNN, the speed-up factor over a gyrokinetic simulation is $4\times 10^5$.
These results demonstrate the effectiveness of our ensemble-based approach in providing accurate and stable predictions across the dataset while maintaining competitive inference performance.

\section{Manual feature engineering}
\label{sec:features}

Complementary to the neural network approach discussed in the previous section, we next present a method to identify patterns in the data using different machine learning algorithms together with manual feature extraction.
While these methods are not yet able to achieve as close a fit to the data as the neural networks, they naturally provide some level of interpretability.
We proceed by first defining a combinatorial procedure to compute a large number of derived features from the raw data that respect the spatial translation invariance.
The Spearman correlation between each feature and the heat flux provides a first measure of each feature's importance.
Then we apply forward sequential feature selection (FSFS) \citep{efroymson1960, whitney1971direct}, where the order in which features are selected provides a second measure of feature importance.
FSFS can be applied using any regression method, and several methods will be compared for the fixed-gradient and varied-gradient data, with the conclusion that there is general agreement.
Finally, a third measure of feature importance is provided by computing Shapley values for the models that result from FSFS.
We will find broad consistency between the three measures of importance.


\subsection{Feature extraction}
\label{sec:feature_extraction}

Since the heat flux from the gyrokinetic model is invariant under periodic translation in $z$, as discussed in section \ref{sec:GKE}, a regression or classification model should preserve this invariance.
However, this invariance would generally be broken if we use the raw features on grid points as inputs to regression or classification methods (other than CNNs) directly.
For example, a regression could in principle find that the heat flux depends specifically on $|\nabla x|^2$ at $z=0.3$, which is not a translation-invariant quantity.
Instead, the inputs to the regression or classification should be translation-invariant features that are derived from the raw $z$-dependent geometric features.
To ensure this invariance in the extracted features, we consider features that are the composition of equivariant operations with translation-invariant reductions.
To explain this method, let $P$ denote the set of periodic real-valued functions on the interval $[0, 1)$, representing the $z$ domain.
Each of the seven original raw features can be considered as an element of the set $P$.
An equivariant operation $E$ is a function from $P\to P$, such that translating the input and then applying $E$ gives the same result as first applying $E$ and then translating.
More formally, let $f$ be any element of $P$, and let $g:P\to P$ indicate a periodic shift by some distance $\Delta$, so $g(f)(x)=f(x + \Delta \mod 1)$.
Then the equivariance of $E$ means that $E(g(f)) = g(E(f))$ for all $f$ and $\Delta$.
A reduction is a map from $P \to \mathbb{R}$. A reduction $A$ is translation-invariant if $A(f(x + \Delta \mod 1)) = A(f(x))$ for all $x$ and any shift $\Delta$.

For the results here, we consider 23 possible translation-invariant reductions: max; min; range (max $-$ min); mean; median; mean of the square; variance; skewness; $L_1$ norm; quantiles 0.1, 0.25, 0.75, and 0.9; count above -2, -1, 0, 1, and 2; the absolute amplitudes of the three longest-wavelength Fourier modes; the expected $k_{||}$; and the largest amplitude $k_{||}$.
The last two quantities are defined as follows: letting $\sum_{j=-N}^N \hat{f}_j \exp(i k_j z)$ denote the Fourier transform of a raw feature $f(z)$, the expected $k_{||}$ is $\sum_{j=1}^N k_j |\hat{f}_j| / \sum_{j=1}^N |\hat{f}_j|$, while the largest amplitude $k_{||}$ is $k_j$ where $j$ is the solution of $\arg\max_j |\hat{f}_j|$.
These last two reductions were included due to the importance of parallel length scales in critical-balance models of turbulence \citep{barnes2011critically}.
We indicate the set of these 23 reductions by $R$.

For the equivariant operations, we consider combinations of unary operations and products.
The unary operations, each an equivariant map $P \to P$, includes 11 operations on an input function $f$:
the identity $f$, $|f|$, $df/dz$, Heaviside$(f)$, Heaviside$(-f)$, ReLU$(f) = \max(f,0)$, ReLU$(-f)$, $1/f$, $f^2$, $fB$, and $f / B$. 
The last two operations, multiplying or dividing by the field strength $B$, were included since the coordinate Jacobian is $1/B$.

We can now state the full set of features that are considered.
Let $F$ denote a set of eight features, given by the original seven raw features together with the local shear $S = (d/dz)(\nabla x \cdot\nabla y / \nabla x \cdot\nabla x)$.
The local shear \citep{greene1968interchange} is added here since it is plausible that it could play a role in determining the turbulence intensity.
Then the set of features $U(F)$ is obtained by applying each possible unary operation to each of the eight elements of $F$.
We let $C(U(F))$ denote the set of all possible pairwise products of functions in $U(F)$, supplemented with $U(F)$.
Then $U(C(U(F)))$ indicates all possible unary operations applied to all elements of $C(U(F))$.
The full set of features is finally $R(U(C(U(F))))$, obtained by applying all of the reductions, for a total of just over 1 million extracted features.
This set includes a few features that are duplicated, improper due to division by 0, or that include unnecessary operations (e.g. ReLU of a positive-definite function), but a suitably large fraction are distinct and finite.
Of course larger feature sets could be considered by including more operations and combinations, but the set here is suitably rich to find accurate regressions of the data.


\subsection{Spearman correlation}
\label{sec:spearman}

One approach to measuring the potential importance of an isolated feature in a regression problem is to compute its Spearman correlation with the target quantity.
Spearman correlation is defined by the Pearson correlation between the sorted rank of the target with the sorted rank of the feature.
The absolute magnitude of the Spearman correlation has the appealing properties of being fast to compute and invariant to any monotonic nonlinear function, e.g. for any sequence of points $\{x_j\}$, the Spearman correlation of $\{x_j\}$ with $\{\exp(x_j)\}$ is 1.
We can evaluate the Spearman correlation between each feature from the previous subsection and the heat flux.
Note that no regression or classification model is used.
The most highly correlated features for the fixed-gradient dataset are listed in table \ref{tab:Spearman}.
All these top features include the factor $\Theta(\mathbf{B}\times\mathbf{\kappa}\cdot\nabla y)$ where $\Theta$ is the Heaviside function.
In our sign convention, this quantity is 1 in regions of bad curvature.
The most correlated features all have a similar form, indicating the heat flux is highest when the flux surface compression $|\nabla x|$ is large in regions of bad curvature.
For the features listed, the sign of the correlations is positive, so greater flux surface compression is associated with a higher heat flux, as expected physically.
The features also include an inverse weighting with $B$, which perhaps reflects the Jacobian $\propto 1/B$.
The exact powers of $|\nabla x|$ and $B$ vary among the top features, but a consistent pattern is evident.

\begin{table}
  \begin{center}
\def~{\hphantom{0}}
  \begin{tabular}{rc}
      \multicolumn{1}{c}{Feature}  & Correlation    \\[3pt]
       variance$(\Theta(\mathbf{B}\times\mathbf{\kappa}\cdot\nabla y) |\nabla x|^2 / B)$   & 0.775 \\
       mean$(\Theta(\mathbf{B}\times\mathbf{\kappa}\cdot\nabla y) |\nabla x|^8 / B^2)$   & 0.774 \\
       mean$(\Theta(\mathbf{B}\times\mathbf{\kappa}\cdot\nabla y) |\nabla x|^4 / B)$  & 0.772 \\
       variance$(\Theta(\mathbf{B}\times\mathbf{\kappa}\cdot\nabla y) |\nabla x|^4 / B)$   & 0.769 \\
       mean$(\Theta(\mathbf{B}\times\mathbf{\kappa}\cdot\nabla y) |\nabla x|^4 / B^2)$ & 0.769 \\
  \end{tabular}
  \caption{Geometric features from section \ref{sec:feature_extraction} with highest absolute magnitude of Spearman correlation to the nonlinear heat flux at fixed temperature and density gradient. Here, $\Theta$ denotes the Heaviside function.}
  \label{tab:Spearman}
  \end{center}
\end{table}


As mentioned in the introduction, these features with highest Spearman correlation to the heat flux are consistent with ideas by \cite{mynick2010optimizing}, \cite{xanthopoulos2014controlling}, \cite{stroteich2022seeking}, and \cite{goodman2024quasi}.
In these earlier works, stellarator configurations were sought with smaller $|\nabla x|$ in regions of bad curvature, motivated by the following physical intuition.
At fixed $dT/dx$, reducing $|\nabla x|$ reduces the real-space temperature gradient $|\nabla T| = (dT/dx)|\nabla x|$.
Reducing $|\nabla T|$ reduces the source of free energy for instabilities and the associated turbulence.
This is particularly true if done in the regions where instabilities and turbulence are localized due to bad curvature.
Positive correlation between $|\nabla x|$ and quasilinear heat flux estimates was also noted by \cite{Jorge2023}.

\changed{
Kendall's $\tau$ is another correlation coefficient with many similarities to Spearman's correlation.
For all results in this paper involving Spearman correlation, essentially the same findings are obtained in comparable computational time if Kendall correlation is used instead, though numerical values of Kendall correlation are smaller.
}

Spearman \changed{and Kendall} correlation cannot account for dependence of the target on multiple features that may interact, so we now proceed to more sophisticated methods.


\subsection{Sequential feature selection}
\label{sec:sfs}

In forward sequential feature selection (FSFS) \citep{efroymson1960, whitney1971direct}, regression or classification is first performed using one feature at a time.
In other words, if there are $n$ features, we fit $n$ distinct models, each with one feature.
Any regression or classification method can be used.
The feature that yields the best model fit to the data is selected to progress to the next step.
Then, the data are fit using $n-1$ independent models that each use two features, the one selected in the first step plus all possible second features.
Of these $n-1$ models, the one with closest fit is selected to progress to the next step, and so forth.
Thus, FSFS results in a parsimonious set of features, which are effectively ranked from most to least important by the order in which they are selected.

For the regression and classification models to use in FSFS, we primarily rely on the gradient boosted decision tree package XGBoost \citep{chen2016xgboost}.
This choice is due to its speed with datasets of this size.
Speed is a priority since $O(10^6)$ independent models must be fit at each step.
Like other decision tree methods, XGBoost fits a piecewise-constant function to the data, with the location and number of discontinuities chosen to balance accuracy of fitting the data against complexity of the surrogate.
At each step of FSFS we use an average score from five-fold cross-validation.
When applying FSFS to the varied-gradient dataset, the temperature and density gradients are also included among the features that can be selected.
Other scalar features such as $\iota$, $d\iota/dx$, $dp/dx$, and $n_{fp}$ can be included, but these features are not selected by the procedure, which makes sense because they do not appear in the gyrokinetic equation and hence have only a more indirect relation to the heat flux.

\begin{figure}
  \centering
   \includegraphics[width=\columnwidth]{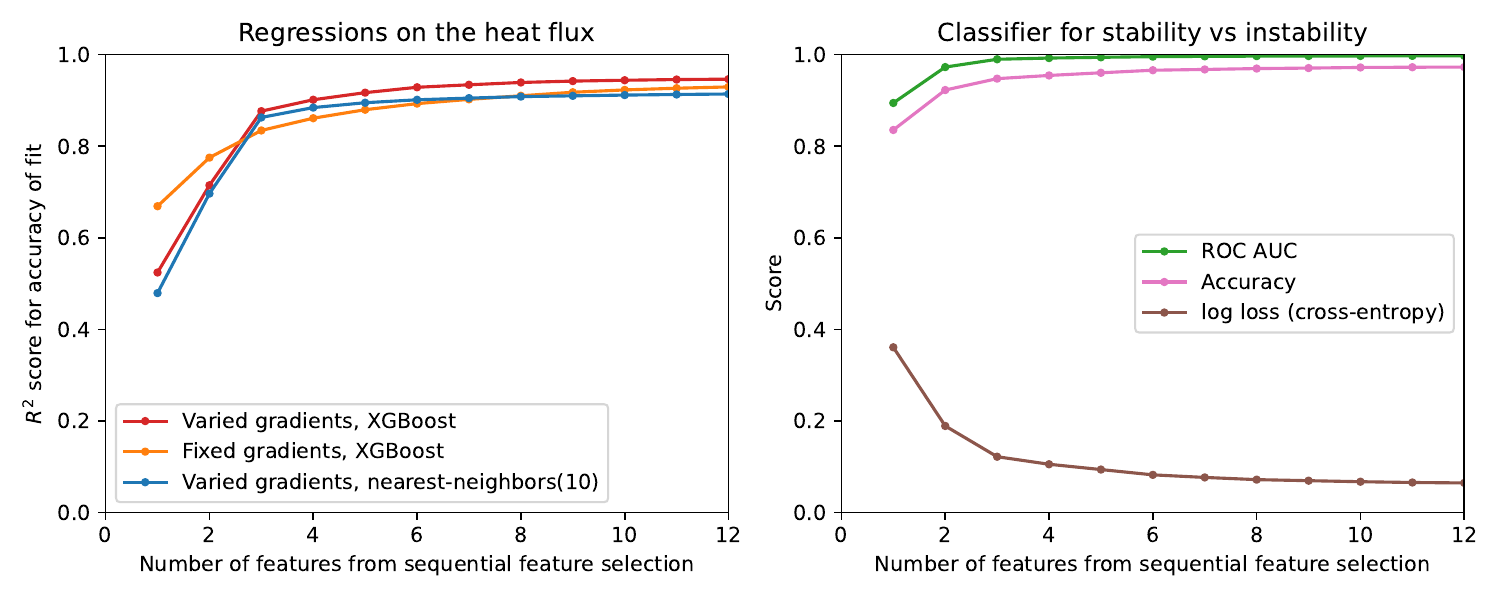}
  \caption{Scores for regression (left) and classification (right) in forward sequential feature selection, showing improvement as the first few features are added.
  Each point shows the mean score on held-out data using 5-fold cross-validation.
  }
\label{fig:sfs}
\end{figure}

Figure \ref{fig:sfs} shows the results of FSFS for regression on both the fixed-gradient and varied-gradient datasets, as well as for stable vs. unstable classification on the varied-gradient dataset.
For regression, we assess the performance using the coefficient of determination $R^2$.
For classification, we choose features in FSFS using the log-loss measure (also known as cross-entropy; lower is better), while reporting also the accuracy and ROC-AUC scores (receiver operating characteristic area under the curve; higher is better) in figure \ref{fig:sfs}.b.
It can be seen that the prediction accuracy rapidly improves with the first few features, then levels off.
With three features, the heat flux for the varied-gradient dataset can be predicted with $R^2=0.88$; adding nine more features results in a marginal improvement to $R^2=0.92$.
Qualitatively similar results are obtained using other regression or classification models.
As an example, figure \ref{fig:sfs} also shows FSFS results for 10-nearest-neighbors regression; the behavior is similar to the XGBoost regression but with slightly lower $R^2$.

\begin{figure}
  \centering
   \includegraphics[width=\columnwidth]{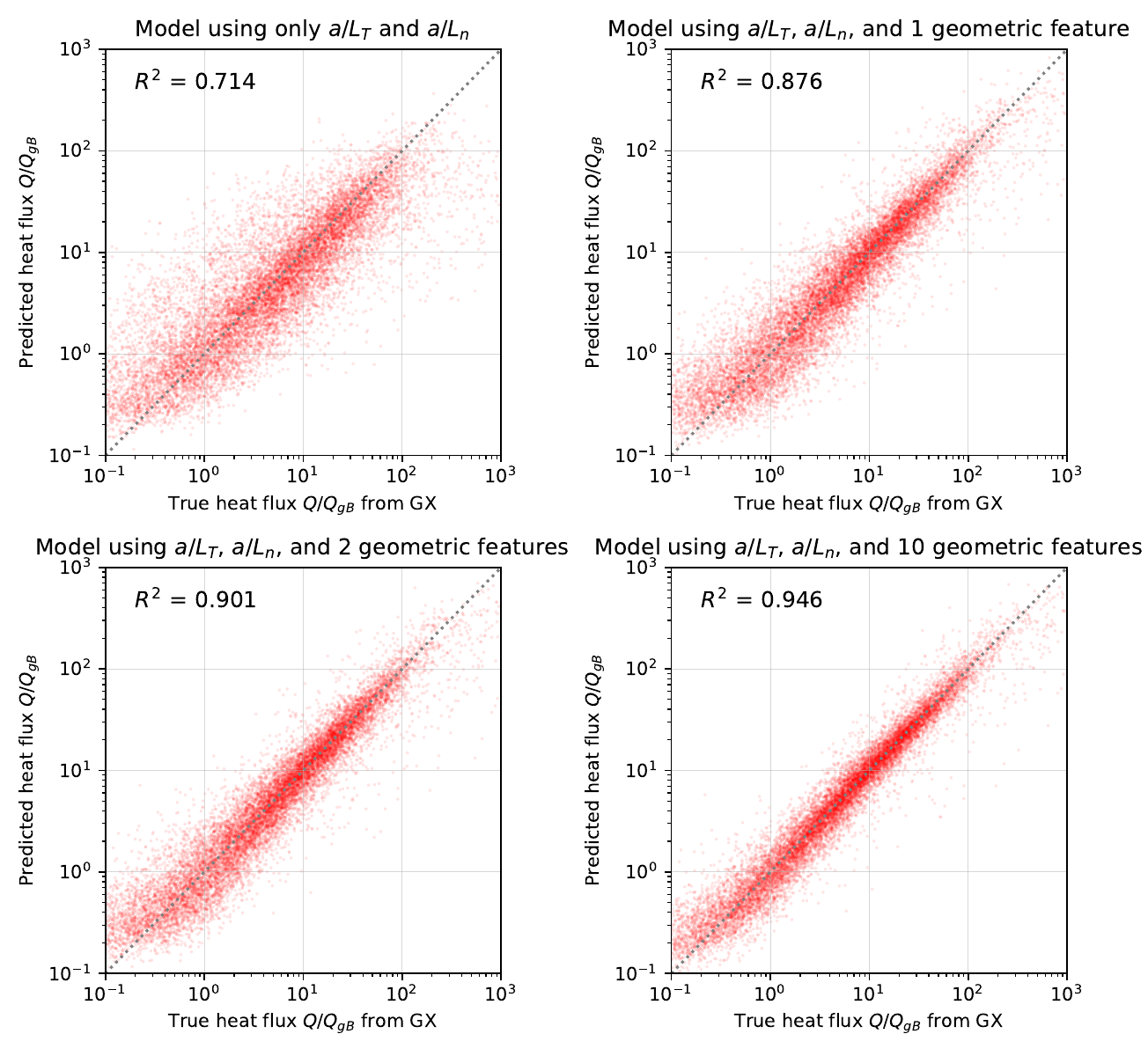}
  \caption{The accuracy of regression models improves to a point as more features are included from sequential feature selection.
  Each panel shows the performance of the XGBoost regression on 20\% held-out test data from the varied-gradient dataset.
  }
\label{fig:sfs_diagonal}
\end{figure}

Another view on FSFS is provided by figure \ref{fig:sfs_diagonal}.
Each panel shows, for a given number of features, how the predictions with that feature subset compare to the actual heat flux.
In each panel, the regression was fit using 80\% of the data, and the figure shows the performance on the held-out 20\% of the data.
A perfect prediction would correspond to all the points lying on the gray dotted line.
As more features are added, the regression models more accurately predict the heat flux.
There is a trade-off in that the models with more features are more complicated and harder to interpret due to possible interactions between the features.

For each of the FSFS curves in figure \ref{fig:sfs}, the first five selected features are listed in table \ref{tab:sfs}.
In the table, absFFTCoeff1 indicates the absolute magnitude of the longest-wavelength (but not constant) Fourier mode in $z$.
In the varied-gradient dataset, for both regression and classification, the first selected feature is the temperature gradient, and the second selected feature is the density gradient.
These findings are consistent with expectations from linear theory that the temperature gradient is the primary drive for the instability, and the density gradient can significantly affect stability.
In both the varied-gradient and fixed-gradient cases, the first geometric feature selected again reflects the flux surface compression in regions of bad curvature.
While the top geometric feature for the classifier involves unfavorable $\nabla B$ drift rather than curvature drift, the feature mean$(\Theta(\mathbf{B}\times\mathbf{\kappa}\cdot\nabla y)|\nabla x|^2/B)$ has almost an identical score, log-loss$=$0.123 as opposed to 0.122, so there is little distinction between the two drifts in this case.
Hence, the results are quite consistent with the Spearman correlation analysis of section \ref{sec:spearman}, and support the physical intuition from \cite{mynick2010optimizing}, \cite{xanthopoulos2014controlling}, \cite{stroteich2022seeking}, and \cite{goodman2024quasi}.

For each of the four tables within table \ref{tab:sfs}, the second geometric feature again includes the flux surface compression $|\nabla x|$, but this time involving the radial rather than binormal component of the $\nabla B$ drift, $\mathbf{B}\times\nabla B\cdot\nabla x$.
Note that there is no distinction between the $\nabla B$ drift and curvature drift in the radial direction: $\mathbf{B}\times\nabla B \cdot\nabla x = B\mathbf{B}\times\mathbf{\kappa}\cdot\nabla x$ for any MHD equilibrium at any $\beta$.
So, the quantity $\mathbf{B}\times\nabla B\cdot\nabla x$ appearing in the second geometric features is the geodesic curvature (times $B|\mathbf{B}\times\nabla x|$).
As mentioned in the introduction, the geodesic curvature has been discussed recently as a correlate of turbulence \citep{xanthopoulos2011zonal, NAKATA2022}, motivated by the fact that geodesic curvature plays a prominent role in the gyrokinetic equation for zonal flow modes \citep{rosenbluth1998poloidal}.
More work would be needed to confirm that the high significance of geodesic curvature in our analysis is due to the effect of zonal flows on the turbulence, but the connection is at least suggestive.

\begin{table}
  \begin{center}
\def~{\hphantom{0}}
  \begin{tabular}{rcrc}
  \multicolumn{2}{c}{\emph{Classification, varied gradients, XGBoost}} &
      \multicolumn{2}{c}{\emph{Regression, varied gradients, XGBoost}} \\[3pt]
      \multicolumn{1}{c}{Feature}  & log-loss   & \multicolumn{1}{c}{Feature}  & $R^2$  \\[3pt]
       $a/L_T$   & 0.361 &
       $a/L_T$   & 0.524 \\
       $a/L_n$   & 0.189 &
       $a/L_n$   & 0.714 \\
       mean$(\Theta(\mathbf{B}\times\nabla B\cdot\nabla y) |\nabla x|^2 / B)$  & 0.122 &
       mean$(\Theta(\mathbf{B}\times\mathbf{\kappa}\cdot\nabla y) |\nabla x|^4 / B^2)$  & 0.876 \\
       mean$(\Theta(-\mathbf{B}\times\nabla B\cdot\nabla x) |\nabla x|^2 B)$   & 0.105 &
       median$((\mathbf{B}\times\nabla B\cdot\nabla x)^2 |\nabla x|^8 / B^8)$   & 0.901 \\
       mean$((\mathbf{B}\times\mathbf{\kappa}\cdot\nabla y) / B)$ & 0.094 &
       absFFTCoeff1$(\mathrm{ReLU}(\mathbf{B}\times\nabla B\cdot\nabla x) / B^5)$ & 0.917 \\
  \end{tabular} \\
  \vspace{0.2in}
    \begin{tabular}{rcrc}
  \multicolumn{2}{c}{\emph{Regression, fixed gradients, XGBoost}} &
      \multicolumn{2}{c}{\emph{Regression, varied gradients, 10NN}} \\[3pt]
      \multicolumn{1}{c}{Feature}  & $R^2$   & \multicolumn{1}{c}{Feature}  & $R^2$   \\[3pt]
       mean$(\Theta(\mathbf{B}\times\mathbf{\kappa}\cdot\nabla y) |\nabla x|^4 / B)$   & 0.669 &
       $a/L_T$   & 0.479 \\
       quantile0.75$((\mathbf{B}\times\nabla B\cdot\nabla x) |\nabla x|^4 / B^5)$   & 0.775 &
       $a/L_n$   & 0.696 \\
       mean$(|\nabla x|^4 / B^6)$  & 0.834 &
       quantile0.9$(\Theta(\mathbf{B}\times\mathbf{\kappa}\cdot\nabla y) |\nabla x|^2 / B)$  & 0.863 \\
       quantile0.9$((\mathbf{B}\times\nabla B\cdot\nabla y) / B^5)$   & 0.861 &
       mean$(\mathrm{ReLU}(\mathbf{B}\times\nabla B\cdot\nabla x) |\nabla x|^2 / B^4)$   & 0.884 \\
       absFFTCoeff1$((\mathbf{B}\times\nabla B\cdot\nabla x) / B^3)$ & 0.879 &
       median$((\mathbf{B}\times\nabla B\cdot\nabla y) |\nabla y|^2 / B^6)$ & 0.895 \\
  \end{tabular}
  \caption{
  First five features from section \ref{sec:feature_extraction} selected with forward sequential feature selection. 
  Results are shown both for classification of stability vs instability, and for regression on the logarithm of the heat flux $Q$.
  Results are also shown for both the gradient-boosted decision tree package XGBoost and for 10-nearest-neighbors (10NN).
  Here, $\Theta$ denotes the Heaviside function.
  }
  \label{tab:sfs}
  \end{center}
\end{table}





At each step of FSFS, there are typically many features which are variations on a theme and which would give nearly the same score, similar to table \ref{tab:Spearman}.
For instance, table \ref{tab:sfs2} shows the top five features in steps three and four of FSFS for regressions on the varied-gradient dataset.
At step 3, the top features are all based on the flux surface compression $|\nabla x|$ in regions of bad curvature, weighted by various powers of the Jacobian.
At step 4, the top features all involve the radial magnetic drift, weighted by the flux surface compression and powers of the Jacobian.
Since the $R^2$ score is so similar between the various options at each step, we should not ascribe too much importance to the details that vary, such as the power of $B$ at step 3.
However aspects that are consistent among the top features, such as the consistent appearance of the radial magnetic drift (i.e. geodesic curvature) at step 4, are more likely to be physically meaningful.

\begin{table}
  \begin{center}
\def~{\hphantom{0}}
  \begin{tabular}{rcrc}
  \multicolumn{2}{c}{\emph{Sequential feature selection step 3}} &
      \multicolumn{2}{c}{\emph{Sequential feature selection step 4}} \\[3pt]
      \multicolumn{1}{c}{Feature}  & $R^2$   & \multicolumn{1}{c}{Feature}  & $R^2$  \\[3pt]
       mean$(\Theta(\mathbf{B}\times\mathbf{\kappa}\cdot\nabla y) |\nabla x|^4 / B^2)$   & 0.876 &
       median$((\mathbf{B}\times\nabla B\cdot\nabla x)^2 |\nabla x|^8 / B^8)$   & 0.901 \\
       mean$(\Theta(\mathbf{B}\times\mathbf{\kappa}\cdot\nabla y) |\nabla x|^4 / B)$   & 0.874 &
       quantile0.75$(\mathrm{ReLU}(-\mathbf{B}\times\nabla B\cdot\nabla x)^2 |\nabla x|^8 / B^6)$   & 0.901 \\
       variance$(\Theta(\mathbf{B}\times\mathbf{\kappa}\cdot\nabla y) |\nabla x|^2 / B)$  & 0.871 &
       median$((\mathbf{B}\times\nabla B\cdot\nabla x)^2 |\nabla x|^8 / B^6)$  & 0.901 \\
       quantile0.9$(\Theta(\mathbf{B}\times\mathbf{\kappa}\cdot\nabla y) |\nabla x|^2 / B)$   & 0.870 &
       median$(|\mathbf{B}\times\nabla B\cdot\nabla x| |\nabla x|^4 / B^4)$   & 0.901 \\
       mean$(\Theta(\mathbf{B}\times\mathbf{\kappa}\cdot\nabla y) |\nabla x|^8 / B^4)$ & 0.869 &
       quantile0.75$(\mathrm{ReLU}(-\mathbf{B}\times\nabla B\cdot\nabla x) |\nabla x|^4 / B^3)$ & 0.901 \\
  \end{tabular} 
  \caption{
  Top-scoring features from steps 3-4 of forward sequential feature selection, for regression on the heat flux using the varied-gradient dataset with XGBoost.
  At each step, there are many features which are variations on a theme that have nearly identical $R^2$ score.
  Here, $\Theta$ denotes the Heaviside function.
  }
  \label{tab:sfs2}
  \end{center}
\end{table}

\begin{figure}
  \centering
   \includegraphics[width=\columnwidth]{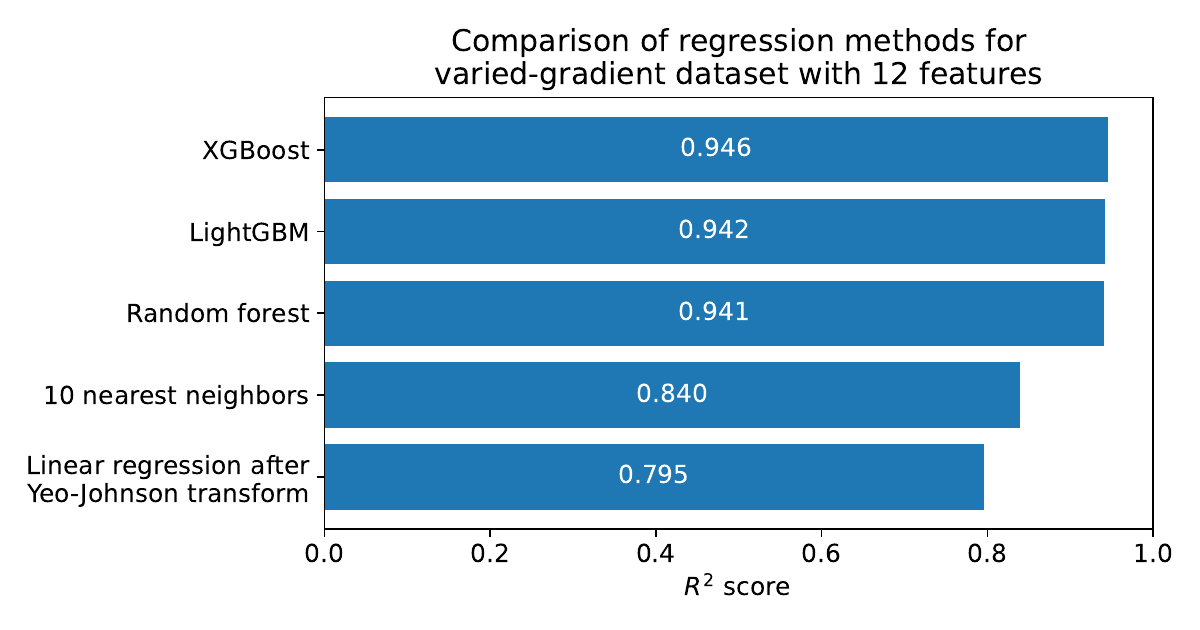}
  \caption{Comparison of regression methods for the heat flux in the varied-gradient dataset.
  In all cases the feature set is $a/L_T$, $a/L_n$, and the top 10 geometric features from FSFS with XGBoost.
  }
\label{fig:other_regression_methods}
\end{figure}

Other regression and classification models besides XGBoost can achieve a reasonable fit to the data.
Figure \ref{fig:other_regression_methods} shows a comparison of several regression methods for the heat flux, all using the same set of 12 features with the varied-gradient dataset ($a/L_T$, $a/L_n$, and the top 10 geometric features from FSFS).
Besides XGBoost and 10-nearest-neighbors (10NN), performance is also shown for three other models.
One is the gradient-boosted decision tree package LightGBM \citep{ke2017lightgbm}.
Another is random forest regression.
Lastly is linear regression after the features are transformed with the Yeo-Johnson power transform \citep{yeo2000new}.
Models other than XGBoost and LightGBM use the implementations from scikit-learn \citep{scikit-learn}.
All models use the default hyperparameters from the relevant python package, and the performance shown is the average score from 5-fold cross-validation.
While the prediction accuracy is highest for the decision-tree methods, it is meaningfully high for the other methods as well.

\subsection{Shapley values}

Another way to measure the importance of the features in regression and classification models is via Shapley values.
A concept originating from game theory, Shapley values are a fair way to divide value among the input features, based on how the model performance degrades when that feature is removed, and considering all subsets of features \citep{shapley1953value,lipovetsky2001analysis}.
Here we specifically use the Shapley Additive exPlanation (SHAP) formulation by \cite{NIPS2017_7062}.
Shapley values naturally provide the average sign of correlation between each feature and the target.
For calculations here we use the SHAP python package \citep{NIPS2017_7062,lundberg2020local2global}, which allows for efficient computations with decision tree models.

\begin{figure}
  \centering
   \includegraphics[width=5in]{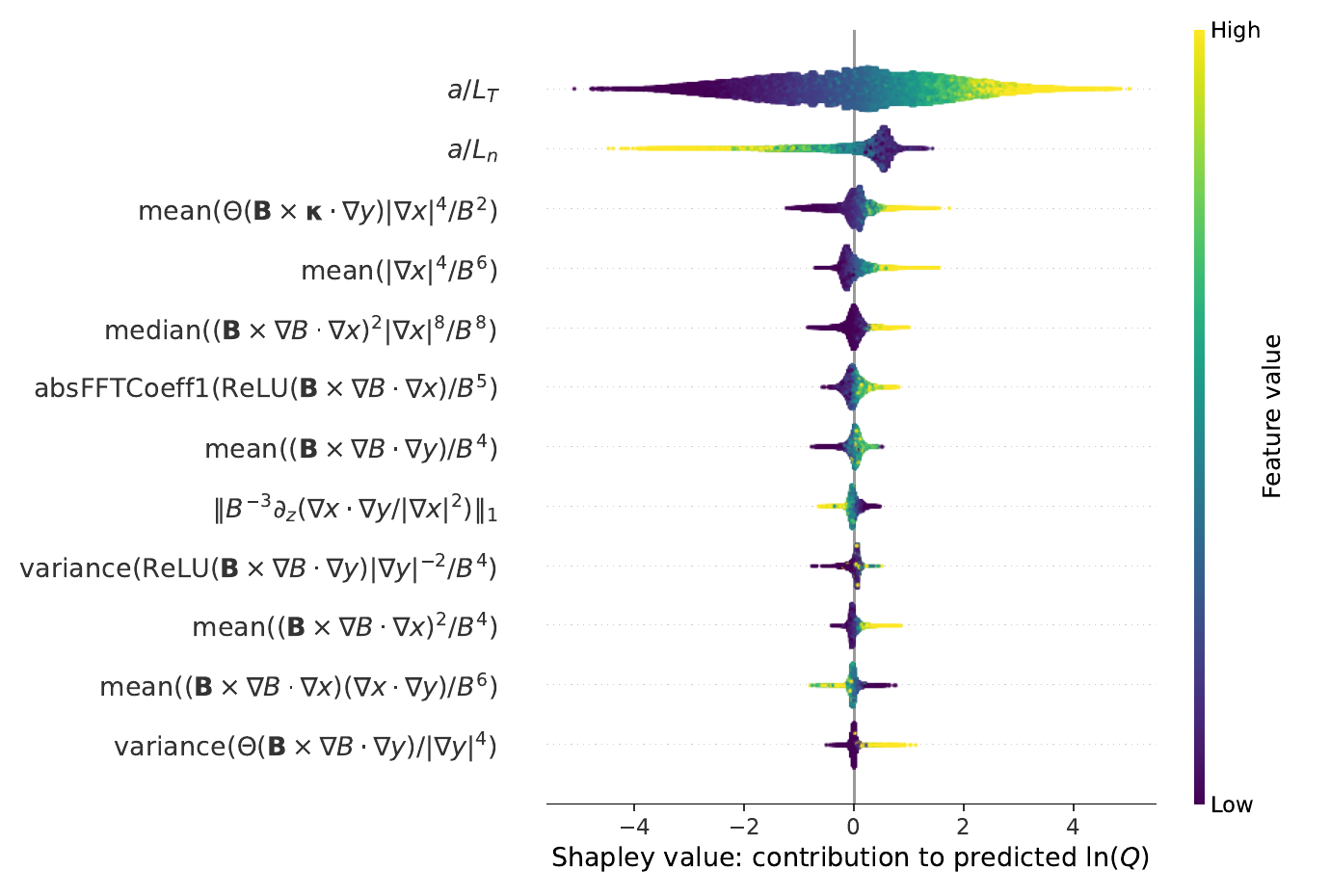}
  \caption{Distribution of Shapley values for regression with the varied-gradient dataset, using an XGBoost fit with top 12 features from FSFS.
  Features are listed in decreasing importance as measured by mean magnitude of the Shapley values.
  }
\label{fig:shapley}
\end{figure}

Figure \ref{fig:shapley} shows the distribution of Shapley values for regression on the varied-gradient dataset, using the first 12 features from FSFS.
Within each row, corresponding to one of the features, a histogram of the Shapley values is shown. 
The horizontal dimension gives the contribution to $\ln Q$ from that feature.
Wider distributions therefore indicate more important features.
The features are sorted in decreasing order of the average absolute Shapley value, i.e. decreasing importance.
Each point is colored by the value of the feature for that flux tube, scaled to the range of that feature over the dataset.
So, if the distribution is purple/dark on the left and yellow/light on the right, larger values of that feature increase the predicted $Q$.
Conversely if the distribution is yellow/light on the left and purple/dark on the right, increasing values of the feature decrease $Q$.

The order of feature importance as measured by Shapley values in figure \ref{fig:shapley} is not exactly the same as the order in which features were selected in FSFS, but the first three features are the same, and the pattern is broadly similar.
Unsurprisingly, the most important feature is $a/L_T$, which is found to increase $Q$.
The next most important feature is $a/L_n$, which decreases $Q$, as expected from linear theory.
Consistent with the findings from Spearman correlation and FSFS, the most important geometric feature is an average of $|\nabla x|$ in regions of bad curvature.
Here, the next most important feature, mean$(|\nabla x|^4/B^6)$, also reflects the average flux surface compression, but now independent of bad curvature.
As the distributions for these two features are blue on the left and red on the right, both of these features increase $Q$, consistent with physical intuition.
The next most important features, median$((\mathbf{B}\times\nabla B\cdot\nabla x)^2|\nabla x|^8/B^8)$ and absFFTCoeff1$($ReLU$(\mathbf{B}\times\nabla B\cdot\nabla x)/B^5)$, both involve the geodesic curvature, consistent with the pattern in table \ref{tab:sfs}.
The colors indicate that larger average magnitude of geodesic curvature increases $Q$, coinciding with the theoretical prediction of \cite{xanthopoulos2011zonal, NAKATA2022}.
Physically, larger geodesic curvature means greater neoclassical damping of zonal flows, hence smaller average zonal flow magnitude, and therefore higher turbulence intensity.

\changed{
The second-most-important geometric feature according to Shapely values, mean$(|\nabla x|^4/B^6)$, was the fifth most important geometric feature according to FSFS order.
Other differences appear in the ranking of feature importance between Shapley values and FSFS after the two geodesic curvature features.
Since Shapely values and FSFS order are different measures of importance, there is no guarantee that they will rank the features in the same order, particularly farther down in the list where the effect sizes become smaller.
}


\subsection{Fine-tuning the top feature}
\label{sec:optimizing_top_feature}

From the subsections above, there is a consistent and robust finding that the most important geometric feature is the flux surface compression in regions of bad curvature.
With this in mind, we can repeat the analyses of the previous subsections with a new set of extracted features that focuses on this general combination of quantities, giving more variations on this theme.
Optimization can also be used to tune constants and exponents appearing within a proposed functional form.
For example, considering features of the form $\mathrm{mean}([\Theta(\mathbf{B}\times\mathbf{\kappa}\cdot\nabla y) + \alpha ] |\nabla x|^\beta / B^\gamma)$, the parameters $\{\alpha,\, \beta,\, \gamma\}$ can be optimized to maximize $R^2$.
Both approaches -- trying a set of variations on a theme and optimization -- result in broadly similar conclusions:
slightly improved prediction of $Q$ is found if a shift is added to the Heaviside function and the exponent on $|\nabla x|$ is modified, yielding the feature
\begin{equation}
    f_Q = \mathrm{mean}([\Theta(\mathbf{B}\times\mathbf{\kappa}\cdot\nabla y) + 0.2] |\nabla x|^3 / B).
    \label{eq:best_feature}
\end{equation}
The shift of 0.2 to the Heaviside function effectively combines the top two geometric features according to the Shapley values in figure \ref{fig:shapley} (though with different powers of $B$.)
For this single feature with the fixed-gradient dataset, the Spearman correlation is 0.788, and XGBoost regression on $\ln Q$ yields $R^2 =$ 0.737.
These values are slightly increased from 0.775 and 0.669 respectively without the shift to the Heaviside function.
For the varied-gradient dataset, using models with the three features $\{ a/L_T, a/L_n, f_Q\}$, regression on $\ln Q$ gives $R^2=0.887$, slightly improved from $R^2=0.876$ from table \ref{tab:sfs}.

\begin{figure}
  \centering
    \includegraphics[width=3.3in]{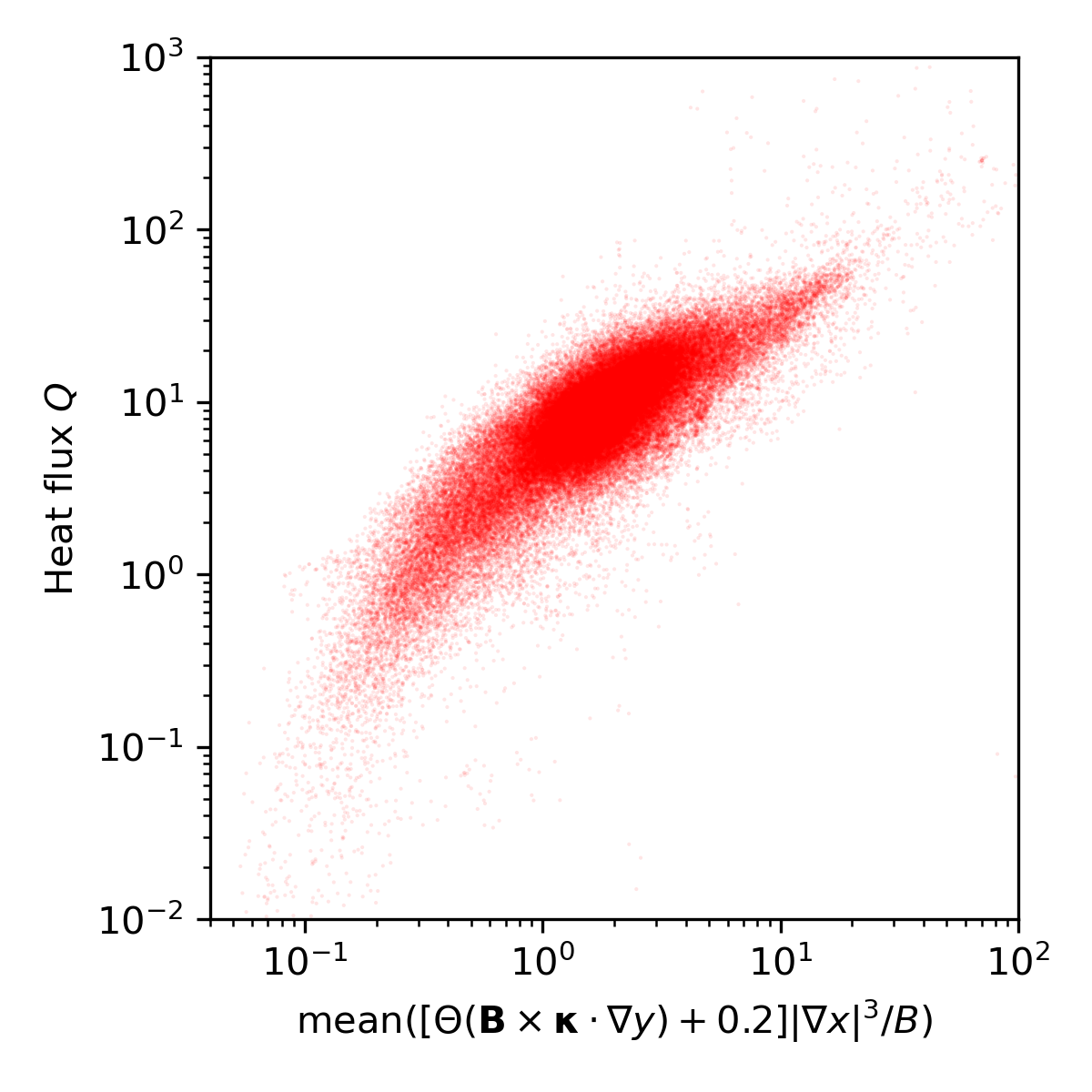}
  \caption{Comparing the true heat flux and the single geometric feature $f_Q$ in eq (\ref{eq:best_feature}) for fixed gradients, it is clear that there is significant correlation.
  No regression model is used here.
  }
\label{fig:single_feature}
\end{figure}

Figure \ref{fig:single_feature} shows a comparison of the single feature $f_Q$ and the GX heat flux for the fixed-gradient dataset.
Note that there is no regression model applied here.
While eq (\ref{eq:best_feature}) does not explain the full variation in the heat flux, it clearly does explain some.
It is possible that with additional work, a single geometric feature of comparable complexity could be found with even higher predictive accuracy.

The set of three features $\{ a/L_T, a/L_n, f_Q\}$ does reasonably well in the classifier for predicting stability: log-loss=0.128, accuracy=0.944, ROC-AUC=0.989.
However, the geometric feature that best predicts the stiffness of the heat flux above marginal stability is not necessarily the best feature for predicting the stability boundary.
Using the same methods described earlier in this subsection, a single geometric feature can be fine-tuned to marginally improve the stability classifier's accuracy.
Again adding a shift to the Heaviside function and modifying the exponents, we arrive at
\begin{equation}
f_\mathrm{stab} = \mathrm{mean}([\Theta(\mathbf{B}\times\nabla B\cdot\nabla y) + 0.4] |\nabla x| / \sqrt{B}).
    \label{eq:best_feature_classifier}
\end{equation}
Using XGBoost with the three features $\{ a/L_T, a/L_n, f_\mathrm{stab}\}$, the classifier achieves log-loss=0.111, accuracy=0.953, and ROC-AUC=0.991.


\section{Testing other proposed surrogates}
\label{sec:other_surrogates}

\begin{figure}
  \centering
    \includegraphics[width=\columnwidth]{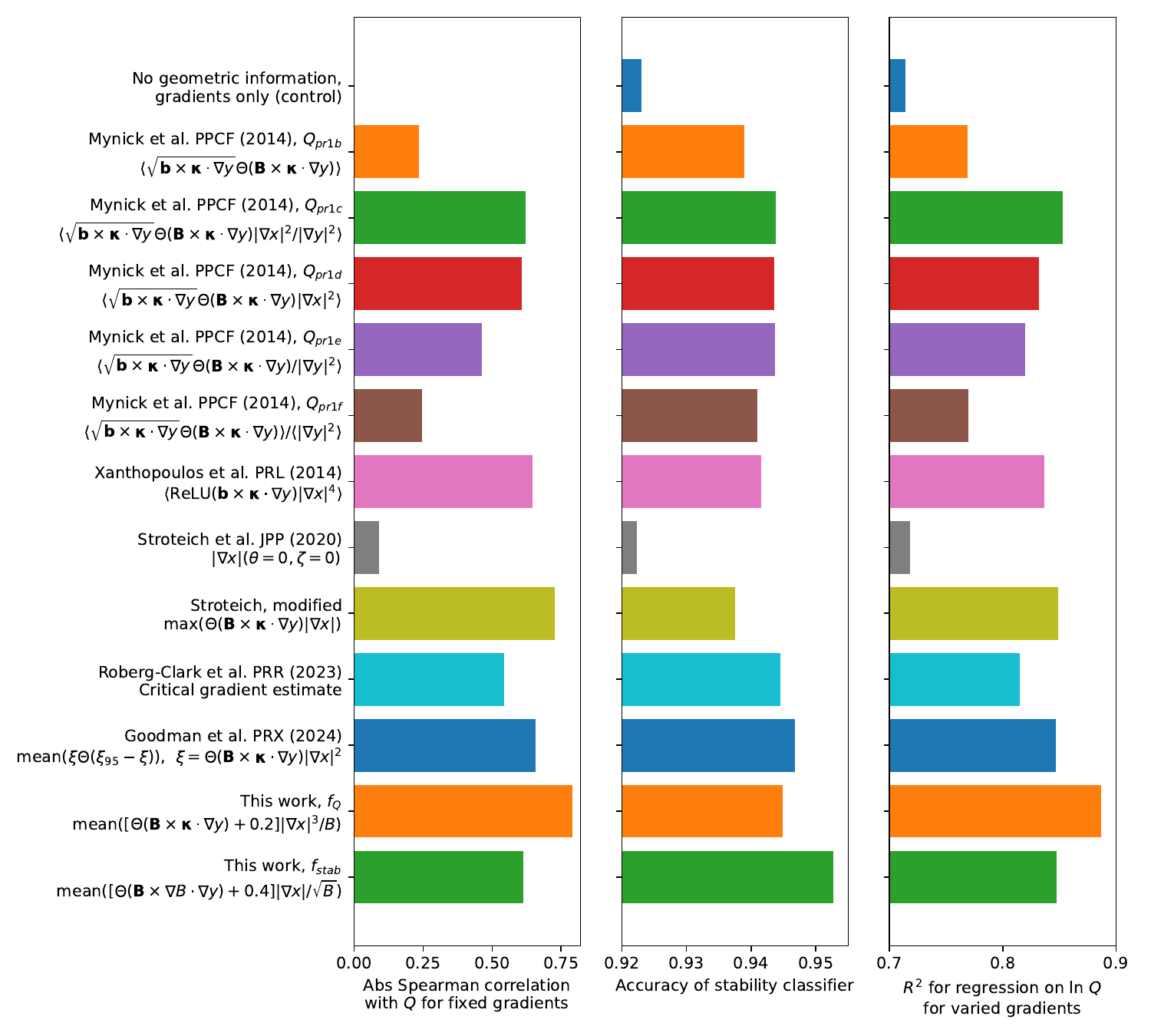}
  \caption{Comparing several proposed ITG objectives using three scores.
  The classification and regression scores are computed using XGBoost with three features: the single geometric feature, $a/L_T$, and $a/L_n$.
  }
\label{fig:previous_proxies}
\end{figure}

A natural application of this dataset is to test objective functions for reduced ITG transport that have been proposed previously.
Several such objectives are compared in figure \ref{fig:previous_proxies}.
Each quantity is rated by three scores: Spearman correlation with $Q$ for the fixed-gradient dataset, accuracy for classification with the varied-gradient dataset, and $R^2$ for regression on the varied-gradient dataset.
For the latter two scores, an XGBoost model is fit using 3 features: the geometric feature in question along with $a/L_T$ and $a/L_n$.
The classification and regression scores are averages from 5-fold cross-validation.
A control is also shown, in which XGBoost is fit using only the two features $\{a/L_T,\, a/L_n\}$ with no geometric features, since the accuracy using only these gradients is already significant.

One set of proposed objectives in figure \ref{fig:previous_proxies} comes from \cite{mynick2014turbulent}: $Q_{pr1b}$-$Q_{pr1f}$.
The definitions can be found in figure \ref{fig:previous_proxies}, in terms of the flux tube average $\langle \ldots \rangle = [\int \ldots dz/B]/\int dz/B$.
In the earlier publication \citep{mynick2014turbulent}, the average was not defined, and it was stated that the component of curvature used was the radial component, but we confirmed with the first two authors that the expressions in figure \ref{fig:previous_proxies} are what was actually used.
Another similar proposed objective in figure \ref{fig:previous_proxies} is the one from \cite{xanthopoulos2014controlling}:
$\langle\mathrm{ReLU}(\mathbf{b}\times\mathbf{\kappa}\cdot\nabla y)|\nabla x|^4\rangle$.
In that paper, the average was not written, and again the component of curvature used was described as radial, but we confirmed with the author that the expression in figure \ref{fig:previous_proxies} is what was actually used.
Our sign convention for $y$ is reversed compared to these earlier papers, so the arguments of Heaviside and ReLU functions have opposite sign.

One more previously proposed objective in figure \ref{fig:previous_proxies} is the one from \cite{stroteich2022seeking}, namely $|\nabla x|$ at a specific point $\mathbf{p}$: the outboard midplane of the taller stellarator-symmetric cross-section.
This quantity is not well correlated with $Q$ in our dataset because 3/4 of the flux tubes in our data do not include this point.
Therefore, we also include a modified objective which is usually similar for flux tubes that include $\mathbf{p}$ but more meaningful in those that do not: the maximum $|\nabla x|$ in regions of bad curvature:
$\mathrm{max}(\Theta(\mathbf{B}\times\mathbf{\kappa}\cdot\nabla y) |\nabla x|)$.

Another feature included in figure \ref{fig:previous_proxies} is the estimated critical gradient from eq (6) of \cite{roberg2023critical}.
We use the curvature drift rather than $\nabla B$ drift to compute this quantity as it results in slightly higher scores, though the differences are small.
A typical flux tube may have multiple intervals of bad curvature, in which case the estimated critical gradient is computed for each interval, and the minimum over all intervals is used as the feature. We note here that the critical gradient estimate is not intended to correlate inversely with heat fluxes, but rather to predict the gradient where heat fluxes become significant. Any inverse correlation of the feature with heat fluxes above the critical gradient, perhaps through the physical connection between $|\nabla \alpha|$ and finite Larmor radius damping of ITG modes, is incidental.

Finally, the comparison in figure \ref{fig:previous_proxies} includes the ITG objective from \cite{goodman2024quasi}.
In that work an integral over a full flux surface was used; here we modify the expression to integrate only over a flux tube.
The specific expression we consider is mean$(\xi \Theta(\xi_{95} - \xi))$, where $\xi = \Theta(\mathbf{B}\times\mathbf{\kappa}\cdot\nabla y)|\nabla x|^2$, and $\xi_{95}$ is the 0.95 quantile of $\xi$ over the flux tube.

Figure \ref{fig:previous_proxies} shows that with the exception of $|\nabla x|(\theta=0,\zeta=0)$, all of these proposed objectives have significant predictive power for both the stability boundary and heat flux.
This is perhaps not surprising as these objectives are similar to each other, all measures of the flux surface compression, bad curvature, or both.
Their performance is nearly as good as for our optimized features $f_Q$ and $f_{stab}$, and is significantly better than the control with no geometric information.


\section{Discussion and future work}
\label{sec:conclusions}

In this work, we have presented a large new dataset of nonlinear gyrokinetic turbulence simulations covering a wide and diverse range of magnetic geometries.
Applied to the dataset, a variety of machine learning methods can accurately predict the nonlinear heat flux, and classify stable versus unstable conditions.
It was important that these classification and regression methods be applied in such a way as to respect the translation-invariance of the gyrokinetic system, as can be done using convolutional neural networks, or translation-invariant reductions of transation-equivariant operations.
Beyond providing these fast surrogates, machine learning methods can also extract insights that can stimulate theory.
Thus, machine learning can be more than a black-box interpolation: it can provide understanding and feed back into more traditional physics calculations.
In order to extract insights in this case, we have demonstrated a variety of methods -- Spearman correlation, sequential feature selection, and Shapley values -- which can measure the importance of geometric features in the data.
While some details differ between the conclusions of these methods, such as the exponent of $B$ in the most important features, certain patterns are quite robust.

Multiple lines of evidence point to the flux surface compression in regions of bad curvature as the most important geometric feature, with more surface compression yielding higher heat flux $Q$.
This feature is identified from its high Spearman correlation, from its early choice in sequential feature selection, and by its large Shapley values.
It arises in FSFS using multiple regression algorithms (decision trees and nearest-neighbors), using both regression and classification, and both with fixed or varied temperature and density gradients.
These findings provide evidence supporting the physical arguments regarding this feature by \cite{mynick2010optimizing}, \cite{xanthopoulos2014controlling}, \cite{stroteich2022seeking}, and \cite{goodman2024quasi}.
It is not obvious that the importance of this feature can be explained purely in terms of linear growth rates, as $|\nabla x|$ does not appear in the linear gyrokinetic equation for \changed{$k_x=0$} modes, which are typically the most unstable.

Another robust finding is that the next most important feature is the geodesic curvature, with larger magnitudes giving higher $Q$.
These conclusions are supported by both FSFS and Shapley values.
Our findings support the discussion of this feature's importance by \cite{xanthopoulos2011zonal} and \cite{NAKATA2022}, and are suggestive of the effect of zonal flow dynamics.
Further analysis of the existing dataset could be done to elucidate the role of zonal flows.

Another theoretical framework for turbulence that has been discussed recently is critical balance \citep{barnes2011critically}, in which the parallel wavenumber $k_{||}$ plays an important role.
Although thousands of features involving $k_{||}$ were included in the set of features that could have been selected, these features were not selected in FSFS, indicating lower importance than the surface compression or geodesic curvature.
It would be valuable to better understand these findings theoretically in the future.

There are many directions in which this research can be extended in the areas of data generation, fitting the data with surrogate models, physics understanding, and applications.
Regarding data generation, the set of equilibria could be expanded with more quasi-isodynamic configurations.
New datasets of nonlinear gyrokinetic simulations could be generated with kinetic electrons and electromagnetic effects.
The physics model would then include additional instabilities such as trapped electron modes and kinetic ballooning modes.
Similar analysis methods to the ones here could then be applied to such data, including also regression on the particle flux.
For both future data and the present data, the feature engineering methods here could be applied with larger sets of possible features beyond the set from section \ref{sec:feature_extraction}, and alternative regression methods could be tried.
Of particular interest would be symbolic regression and Kolmogorov Arnold networks \citep{liu2024kan} due to their advantages for interpretability.
It would also be valuable to use saliency maps and related methods to understand the features learned by the neural networks.
If these saliency maps can be understood, the results may suggest new features that could be checked directly for correlation with the true heat flux and to include in the FSFS.

In the area of physics understanding, researchers can aim to  derive relationships between the top geometric features here and the nonlinear heat flux using traditional analytic methods.
This should be done first for the most important feature (and variations thereof) related to flux surface compression, but could also be attempted for the next most important features.
Other quantities inspired by plasma theory could be checked for correlations against the heat flux, and added to the menu of possible features for sequential feature selection.
It would be natural for instance to check quantities from linear zonal flow dynamics in this way.

The surrogate models developed in this paper could be applied in multiple ways.
One application is for predicting the radial profiles of temperature, using the surrogate as a fast model for the gradient-flux relationship inside a solver of the transport equations.
(Prediction of the density profile would require a dataset of nonlinear simulations with kinetic electrons.)
Such profile prediction using surrogate models is already available for tokamaks \citep{citrin2015real,meneghini2017self}, but this could now be extended to stellarators.
The other evident application would be for turbulence optimization of stellarators.
While optimization of a geometric feature similar to the top feature here is already being done \citep{goodman2024quasi}, the results of this paper allow several improvements.
First, this turbulence objective can now be justified through validation on this data, and modified slightly as in section \ref{sec:optimizing_top_feature} for better correlation to nonlinear simulations.
Second, the surrogates here provide better correlation to the true heat flux by incorporating more geometric information, either via multiple features in the FSFS method, or through holistic use of all the geometric information in the neural networks.
Lastly, since the surrogates here could allow information from all flux surfaces and the gradients to be combined in a transport solver to rapidly predict the fusion power, this enables the fusion power to be used directly as an optimization objective.



\section*{Acknowledgements}
We gratefully acknowledge suggestions related to this work from
Ian Abel,
Gonçalo Abreu,
David Bindel,
Bill Dorland,
Alan Goodman,
Greg Hammett,
Siena Hurwitz,
Byoungchan Jang,
Rogerio Jorge,
Alan Kaptanoglu,
Emily Kendall,
Ralf Mackenbach,
\changed{Maikel Morren,}
Harry Mynick,
Eduardo Rodriguez,
and
Pavlos Xanthopoulos.

\section*{Funding}
This work was supported by the US Department of Energy under DE-AC02-09CH11466 (High-fidelity Digital Models for Fusion Pilot Plant Design, StellFoundry).
R. C. was supported by Greg Hammett's DOE Distinguished Scientist Fellow award.
This research used resources of the National Energy Research Scientific Computing Center (NERSC), a Department of Energy Office of Science User Facility using NERSC awards FES-mp217 and FES-m4505 for 2024.

\section*{Declaration of interests}
M. L. is a consultant for Type One Energy Group.

\section*{Data availability statement}
Data associated with this study can be downloaded from \url{https://doi.org/10.5281/zenodo.14867776} \citep{zenodo}.


\appendix

\section{Details of equilibrium generation}
\label{app:equilibrium_details}

As discussed in section \ref{sec:equilibrium_generation}, the equilibria in this study consist of three classes.
Here we give additional details of how the equilibria in each class were generated.
In all cases, the equilibria were computed using  DESC \citep{dudt2020desc} release v0.12.0.

For the group of rotating-ellipse equilibria, the number of field periods was sampled randomly from  the interval $[2, 8]$, the aspect ratio was chosen randomly from [6, 10], and the elongation (ratio of major to minor axis of the cross-section in a constant-$\phi$ plane) was chosen randomly from $[1, 4]$.
For half of the rotating-ellipse configurations, the ellipses were centered on a constant-major-radius circle.
For the other half, the ellipses were centered on a curve with torsion, by setting the $m=0$, $n=n_{fp}$ Fourier mode of $R$ and $Z$ on the boundary to a random number between 0 and the minor radius, with sign chosen to increase iota.
A pressure profile with fixed shape $p(s) = 1-1.8s+0.8 s^2$ is chosen, where $s$ is the normalized toroidal flux, reflecting a plausible level of peaking, with random magnitude chosen for a uniform distribution of volume-averaged $\beta \in [0, 0.05]$.
Equilibrium calculations were then run, assuming no toroidal current for simplicity.
Configurations with $|\iota| < 0.2$ were dropped.

Another group of equilibria were derived from the QUASR database \citep{giuliani2024direct, giuliani2024comprehensive}.
This database includes coils in addition to plasma shapes, and many entries in QUASR differ primarily in coil geometry while having similar plasma parameters.
Also, some configurations in QUASR have much better quasisymmetry than others, and we wish to focus on the ones with better quasisymmetry (since the other equilibrium groups in our study contain many non-quasisymmetric geometries.)
Therefore a subset of QUASR was selected as follows.
First, all configurations with high quasisymmetry error, $\iota< 0.2$, or aspect ratio $> 10$ were excluded.
Then for each symmetry class (QA vs QH) and $n_{fp}$, for each interval in aspect ratio (1-2, 2-3, 3-4, .., 9-10) and $\iota$, the two configurations in QUASR with lowest QS error were chosen.
DESC was run for these boundary shapes assuming a vacuum field, and also with the aforementioned pressure profile for multiple pressure magnitudes spanning $\langle\beta\rangle \in [0, 5\%]$.
For some strongly shaped geometries or high $\langle\beta\rangle$ values, the resulting force residual was high, indicating the equilibrium was not well converged, so these cases were dropped.

The third group of equilibria were generated by randomly sampling Fourier modes from distributions that have been fit to a dataset of previous stellarator shapes.
Consider the common representation of toroidal boundary shapes as a double Fourier series:
\begin{equation}
R(\theta,\phi)=\sum_{m,n} R_{m,n} \cos(m\theta-\nfp n\phi),
\hspace{0.25in}
Z(\theta,\phi)=\sum_{m,n} Z_{m,n} \sin(m\theta-\nfp n\phi),
\end{equation}
where $(R,\phi,Z)$ are cylindrical coordinates,  $\nfp$ is the number of field periods, $\theta$ is some poloidal angle, and stellarator symmetry has been assumed.
The sums are considered to include only non-negative $m$, with both positive and negative $n$ for $m>0$, but only non-negative $n$ for $m=0$.
To randomly generate boundary shapes, we sample the $R_{m,n}$ and $Z_{m,n}$ coefficients from independent normal distributions for each $(m,n)$ (and independent for $R$ and $Z$).
The mean and standard deviation of each distribution are taken from the dataset of 44 stellarators collected by \cite{kappel2024magnetic}.
This set includes both built experiments (W7-X, LHD, HSX, CFQS, TJ-II, etc.) and theoretical configurations.
Before extracting the sample mean and standard deviation, all configurations were scaled to the same minor radius.
To further standardize the data, $\phi \to -\phi$ reflections were applied to configurations with $\iota < 0$ so all configurations have rotational transform of the same sign, and toroidal rotations by half a field period were applied to any configurations in which the cross-section at $\phi=\pi/\nfp$ is not as tall as at $\phi=0$.
After this data cleaning, the sample mean and sample standard deviation are computed over the 44 configurations for each $R_{m,n}$ and $Z_{m,n}$.
Configurations with all values of $\nfp$ are included together in this calculation.
Only modes with $m \le 4$ and $|n| \le 4$ are considered, since some theoretical stellarators in Kappel's dataset resulting from optimization do not include boundary modes with higher mode number.
The resulting mean and standard deviation values are used to define normal distributions which can then be sampled.

To generate a new random equilibrium, new values of $R_{m,n}$ and $Z_{m,n}$ are first sampled from the distributions determined above, after which $R_{0,0}$ is computed by root-finding to give the desired aspect ratio.
To ensure that the same gyro-Bohm normalization is used in every turbulence simulation, each configuration is scaled slightly to the same minor radius, and the toroidal flux is set to the same value.
As with the other equilibrium groups, a pressure profile with fixed shape $p(s) = 1-1.8s+0.8 s^2$ is chosen, with random magnitude chosen for a uniform distribution of volume-averaged $\beta \in [0, 0.05]$.
The current profile is taken to be zero for simplicity.
Next, a fast and low-resolution equilibrium calculation is run using the code VMEC \citep{hirshman1983steepest}.
If this calculation does not converge to a threshold value of force residual in a given number of iterations, the configuration is rejected, ensuring that self-intersecting boundaries are excluded quickly.

A potential issue with the procedure above is that most resulting configurations in the third group have larger values of mirror ratio $B_{max} / B_{min}$ than typical stellarators.
For this reason, configurations with $B_{max} / B_{min} > 5$ are immediately rejected.
Also, a subset of the random configurations is selected to bias the distribution towards smaller mirror ratios.
For the selected configurations, higher-resolution equilibrium calculations are then run using DESC.

The final set of 23,577 equilibria included
3,200 rotating-ellipse configurations centered on a circle,
3,200 rotating-ellipse configurations centered on a curve with torsion,
413 QUASR vacuum configurations,
3,964 QUASR finite-beta configurations,
and 3,200 random boundaries for each of four values of $n_{fp}$.
More flux tubes were drawn from the QUASR vacuum fields to increase the representation of this class.
The final set of 100,705 flux tubes included
12,795 tubes from rotating-ellipse configurations centered on a circle,
12,791 tubes from rotating-ellipse configurations centered on a curve with torsion,
8,235 tubes from QUASR vacuum configurations,
15,809 tubes from QUASR finite-beta configurations,
and 51,075 tubes from random-boundary equilibria.



\section{Details of turbulence simulations}
\label{app:turbulence_details}

Here, more details are given of the nonlinear turbulence simulations.
Calculations were performed using the version of GX from git commit b88d763.

For the dataset with varied gradients, the gradients were sampled randomly in each simulation using the following procedure.
Since $a/L_{Ti}$ typically increases with radius, we choose it to be $\rho$ times a random number with mean 4 and standard deviation 3, resampling the latter until the result is $\ge 1.5$.
Similarly, $a/L_n$ is chosen to be $\rho$ times a random number with mean 1 and standard deviation 2, resampling until the number is $\ge -0.5$.
This procedure results in 70\% of the simulations yielding instability, 30\% stable.

In all simulations, physical collisions were included with magnitude $\nu_{ii} a_\mathrm{minor} / v_{i}=0.01$, and we take $T_i / T_e = 1$.
\changed{
The transient behavior for $t v_{i} / a_\mathrm{minor}<150$ was ignored, and a mean of the heat flux was computed over the remaining simulation time.
}

\begin{figure}
  \centering
  \includegraphics[width=3in]{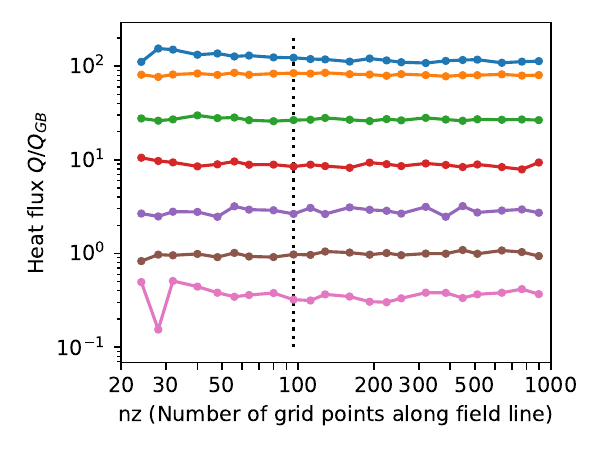}
  \caption{
  \changed{
  Dependence of $Q$ on the resolution parameter \texttt{nz} for a selection of flux tubes with associated gradients from the varied-gradient dataset.
  The seven tubes, shown by different colors, all have the highest value of $n_{fp}$ in the dataset (8) to give short scales in $z$.
  The vertical dotted line shows the value of \texttt{nz} used for the main dataset (96).
Variation of $Q$ with increasing resolution is small compared to the differences between geometries and gradients, indicating sufficient convergence.
  }
  }
\label{fig:nz_convergence}
\end{figure}

\begin{figure}
  \centering
  \includegraphics[width=4in]{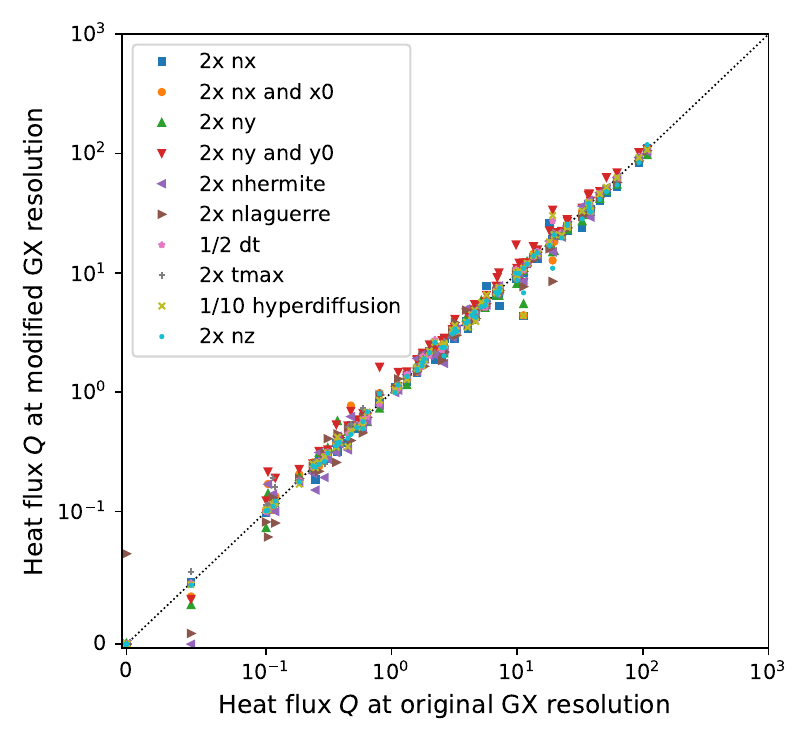}
  \caption{
  \changed{
  Evidence of sufficient convergence with respect to numerical resolution parameters.
  For 100 randomly sampled entries in the varied-gradient dataset, every resolution parameter is varied by a factor of 2 or 10.
  The box sizes in $x$ and $y$ are denoted $x0$ and $y0$ in the legend.
  }
  }
\label{fig:convergence100}
\end{figure}

\changed{
Resolution parameters for the simulations were as follows:
box size \texttt{x0}=\texttt{y0}=10 * 2$\pi$,
perpendicular spatial resolution \texttt{nx}=\texttt{ny}=64,
(so the dealiased $k_x$ and $k_y$ grids were $2.1, -2.0, \ldots,-0.1, 0, 0.1, \ldots, 2.0, 2.1$),
number of grid points along the field line \texttt{nz}=96,
parallel velocity resolution \texttt{nhermite}=8,
perpendicular velocity resolution \texttt{nlaguerre}=4,
simulation time $t_\mathrm{max} v_{i} / a_\mathrm{minor}=800$,
and time step given by 0.9 times the Courant–Friedrichs–Lewy condition.
To arrive at these values, first each of these parameters was scanned individually for a selection of flux tubes and gradients.
An example is shown in figure \ref{fig:nz_convergence}, in which \texttt{nz} is varied for several configurations with $n_{fp}=8$, the highest value of $n_{fp}$ included in the dataset, resulting in the shortest parallel length scales among the flux tubes in the collection.
For each parameter, a value was adopted at which $Q$ had approximately reached an asymptote.
For changes to the box size in $x$ or $y$, \texttt{nx} or \texttt{ny} was varied proportionally to keep the highest wavenumber fixed.
Flux tube length was not considered to be a resolution parameter like the others mentioned above, since longer tubes do sample different regions of the flux surface geometry, so changes to $Q$ are physical rather than numerical.
Then to confirm that the selected resolution parameters would be sufficient for most simulations in the dataset, 100 random flux tubes and gradients were selected from the varied-gradient data.
Every resolution parameter was varied by a factor of two or more for each flux tube and GX was re-run.
The results are displayed in figure \ref{fig:convergence100}.
The coefficient of determination between the original-resolution data and modified-resolution data is $R^2=0.993$ for the unstable cases, or including both stable and unstable cases $R^2=0.995$  using eq (\ref{eq:ln_1_plus_Q}).
The accuracy score for predicting the stability at high resolution from the stability at standard resolution is 0.994.
These values are greater than the values for any of the surrogate fits discussed in this paper, meaning scatter due to discretization error is smaller than imperfection in the fits.
We therefore deem the resolution parameters to be adequate.
}

\begin{figure}
  \centering
  \includegraphics[width=3in]{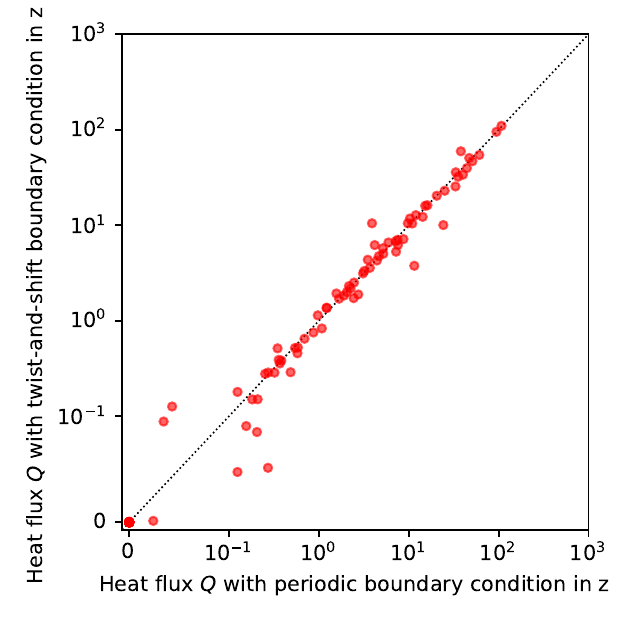}
  \caption{
  \changed{
Comparison of boundary conditions in $z$ for 100 randomly sampled flux tubes from the varied-gradient dataset.
The heat flux is insensitive to the choice of boundary condition if the number of Fourier modes in $x$ is increased in each twist-and-shift calculation to match the same maximum $k_x$ as the periodic calculations, as is done here.
  }
  }
\label{fig:z_boundary_condition}
\end{figure}

Periodic boundary conditions were employed in all three spatial coordinates.
\changed{
For the $z$ coordinate, the twist-and-shift boundary condition \citep{beer1995field} and its generalization to stellarators \citep{martin2018parallel}, while well-motivated physically, are inconvenient when the integrated local magnetic shear is small.
In this case the box size in $x$ is required to be very large, necessitating large \texttt{nx} in order to resolve adequately high $k_x$, increasing computation time.
Were we to use the alternative boundary conditions by \cite{martin2018parallel} that constrain the tube length to certain specific values, it would no longer be possible to use the same tube length for all configurations, complicating the analysis.
\cite{martin2018parallel} found that the heat flux is insensitive to the choice of boundary conditions as long as enough Fourier modes in $x$ are included to resolve sufficiently high $k_x$.
We find the same to be true for the simulations here.
Figure \ref{fig:z_boundary_condition} shows a comparison of periodic vs. twist-and-shift boundary conditions (using eq (24)-(25) of \cite{ martin2018parallel}) for a random sample of 100 flux tubes and associated gradients from the varied-gradient dataset.
For the twist-and-shift calculations, the box size in $x$ is set by the quantization condition, and \texttt{nx} is increased as needed for each tube to match the same maximum $k_x$ as in the periodic case ($k_x = 2.1$).
The heat fluxes for the two choices of boundary conditions are highly correlated.
The $R^2$ for predicting $\ln(Q_{linked})$ from $\ln(Q_{periodic})$ is 0.97, and the accuracy score for predicting stability for the twist-and-shift case based on stability for the periodic case is 0.95.
These values are deemed sufficiently high that the periodic boundary condition was adopted for the study, given the computational savings it allows.
}

At \changed{the resolution parameters given above}, the mean simulation wallclock time was $\sim 8$ minutes on one Nvidia A100 GPU.
The $\sim 2\times 10^5$ nonlinear simulations took $<7000$ node-hours, equivalent to $<28,000$ gpu-hours (4 gpus/node).

In terms of normalized variables, the heat flux returned directly by GX is $Q / \langle |\nabla\rho|\rangle$ where $Q=\langle \int d^3v\, f (v^2/2) \mathbf{v}_d\cdot\nabla\rho \rangle$ and $\langle \ldots \rangle=(\int dz/B)^{-1} \int dz (\ldots)/B$.
For results in this paper, we multiply through by $\langle |\nabla\rho|\rangle$ to focus on $Q$ itself.

Simulations with $Q>10^3$ are dropped from the dataset, since longer simulation times and larger box sizes are likely needed for adequate resolution of these cases.

\bibliographystyle{jpp}

\bibliography{turbulence_ml}

\end{document}